\documentstyle[eqsecnum,preprint,aps,epsfig,floats]{revtex}
\tighten

\newcommand{\Beq}{\begin{equation}}
\newcommand{\Eeq}{\end{equation}}
\newcommand{\beq}{\begin{displaymath}}
\newcommand{\eeq}{\end{displaymath}}
\newcommand{\Beqa}{\begin{eqnarray}}
\newcommand{\Eeqa}{\end{eqnarray}}
\newcommand{\beqa}{\begin{eqnarray*}}
\newcommand{\eeqa}{\end{eqnarray*}}
\newcommand{\Bml}{\begin{mathletters}}
\newcommand{\Eml}{\end{mathletters}}

\begin{document}

\preprint{IRB-TH-2/98}

\draft

\title{Perturbative approach to the penguin-induced $B \rightarrow 
       \pi \phi$ decay}

\author{Bla\v zenka Meli\'{c}\thanks{melic@thphys.irb.hr}}

\address{Theoretical Physics Division, Rudjer Bo\v{s}kovi\'{c} Institute, \\
        P.O. Box 1016, HR-10001 Zagreb, Croatia}

\date{\today}

\maketitle

\begin{abstract}
Using a modified perturbative approach that includes the Sudakov resummation
and transverse degrees of freedom we analyze the penguin-induced $B^{-}
\rightarrow \pi^{-}\phi$ decay by applying the next-to-leading order 
effective weak Hamiltonian. The modified perturbative method enables 
us to include nonfactorizable
contributions and to control virtual momenta appearing in the process.
Besides, we apply the three-scale factorization theorem for nonleptonic 
processes that offers the possibility of having the scale-independent 
product of short- and long-distance parts in the amplitude of the 
weak Hamiltonian. 
The calculation supports the results obtained in the BSW factorization 
approach, illustrating the electroweak penguin dominance and the 
branching ratio of order ${\cal O}(10^{-8})$. 
However, the estimated prediction of $16\%$ for the CP asymmetry is much 
larger than that obtained in the factorization approach. 
\end{abstract}

\vskip 0.5cm

\narrowtext

\section{Introduction}

Among a variety of heavy-meson decaying channels, exclusive 
two-body nonleptonic decays are theoretically the most 
challenging ones, 
owing to the phenomenon of hadronization and the effects of final-state 
interactions. On the other hand, they present the most promising 
way to detect CP 
violation in the heavy-meson sector and to explore the CKM mixing matrix 
elements.

The mechanism of CP violation can be investigated directly in the 
charged sector of $B$ mesons by measuring CP asymmetry. 
CP asymmetry is defined as a relative difference between of the decay rates of 
the $B$ meson and its CP-conjugated state, i.e.,
\Beq
a_{\rm CP} = \frac{\Gamma (B^{-}\rightarrow f) - \Gamma (B^{+}\rightarrow 
\overline {f})} {\Gamma (B^{-}\rightarrow f) + 
\Gamma (B^{+}\rightarrow \overline{f})} \; .
\label{eq:aCP}
\Eeq
Nonvanishing CP asymmetries appear through the interference between amplitudes 
with different weak CP-violating phases and different CP-conserving 
strong phases coming 
from the final-state strong interactions different from zero.

Nowadays, experimental facilities offer a possibility of searching for CP asymmetries 
in penguin-induced nonleptonic decays, the very promising decays to detect 
direct CP violation. Such decays have small branching ratios (BR), but satisfy 
both requirements for CP-violating asymmetry, owing to the fact that 
penguins are loop diagrams with different quark generations contributing with 
different weak CP-phases from the CKM matrix and that 
final-state strong interaction phases emerge from the  absorptive part of 
penguin amplitudes. This mechanism of generating CP asymmetries in 
decays that involve penguins was first considered by Bander, Silverman, and 
Sony \cite{BSS}.

In this paper we discuss the pure penguin-induced 
$B^{-} \rightarrow \pi^{-}\phi$ decay governed by the heavy-quark 
$b \rightarrow d s \overline{s}$ decay. Performing a consistent $1/N_c$ expansion 
($N_c$, quark color number) of QCD-penguin amplitudes, 
one can show 
that in this process the QCD-penguin contribution should 
be suppressed and the dominant contribution comes from the 
electroweak (EW)-penguin 
operators \cite{Flrev}. The $B^{-} \rightarrow \pi^{-}\phi$ process 
has already been considered 
by many authors \cite{Fl}-\cite{ALI}
 within the Bauer-Stech-Wirbel (BSW) factorization approach
\cite{BSW}. This method for reducing the hadronic matrix element of four-quark 
operators to the product of two current-matrix elements cannot account 
for QCD interactions between the currents, except by parametrizing them by a 
phenomenological parameter in the generalized factorization approach \cite{NS}. 
In general, momenta of virtual gluons or 
photons in a process, appearing explicitly in the penguin matrix elements 
after factorization 
have to be considered free parameters. CP asymmetry depends strongly 
on these parameters and the predictive power of calculations performed within 
the factorization prescription is greatly reduced.

Our aim is to investigate the $B^{-} \rightarrow \pi^{-}\phi$ decay in the 
modified perturbative approach. Perturbative calculations of exclusive 
$B$ decays were carried out by different authors \cite{SHBetc}, all of them 
following the framework for analyzing exclusive decays in the perturbative 
QCD (pQCD) approach developed by Brodsky and Lepage, and other authors \cite{BLetc}. 
In the perturbative approach, exclusive amplitudes involving large momentum 
transfer factorize into a convolution of a process-independent and 
perturbatively incalculable distribution amplitudes (hadronic wave functions), 
one for each hadron involved into the decay, with a process-dependent and 
perturbatively calculable hard scattering amplitude of valence partons. 

The applicability of such a pQCD framework to exclusive decays was widely discussed 
\cite{Ra}, \cite{IsLl} owing to the concern about the possible uncontrollable 
nonperturbative (end-piont region) contributions and the problem was solved 
in a modified perturbative approach proposed by Li and Sterman \cite{LiSt}. 

Besides offering a reliable perturbative calculation, the modified perturbative 
approach offers a possibility of going beyond 
the factorization approximation in the calculation of four-quark matrix 
elements. It also enables us to assign the 
process-dependent virtual momenta $q^2$ in the loop matrix elements 
and to fold them with their distribution 
in a particular decay. In this way, the 
uncertainties in CP asymmetry, which are due to some ad hoc quark model 
values of $q^2$ as applied in the factorization approach, 
do not appear.

The purpose of this paper is to present a complete calculation of factorizable 
and nonfactorizable contributions in the penguin-induced 
$B^{-} \rightarrow \pi^{-}\phi$ decay up to order $\cal O (\alpha_{\rm s} 
\alpha_{\rm em})$, 
testing the results on various CKM mixing matrix parameters. 
Especially, we wish to examine 
nonfactorizable contributions from the QCD-penguin operators and 
assign their role in the EW-penguin dominated processes, such as 
$B^{-} \rightarrow \pi^{-}\phi$.

The plan of the paper is as follows. In Sec. II we introduce the method 
of calculation based on the next-to-leading order (NLO) effective weak 
Hamiltonian, and the modified perturbative approach. A detailed analysis 
of the $B^{-} \rightarrow \pi^{-}\phi$ process is presented in Sec. III. 
A discussion of mesonic wave functions and the Sudakov form factors 
is given in Sec. IV and the selection of proper mesonic wave functions 
is made. In Sec. V we present 
our numerical results for the branching ratio and CP asymmetry, comparing 
them with those obtained from the factorization approach and examine 
the dependence of the results on the choice of CKM parameters.
 Concluding remarks are given in 
Sec. VI.

\section{Perturbative model for calculating the $B^{-} \rightarrow \pi^{-}\phi$
decay}

The nonleptonic $B^{-} \rightarrow \pi^{-}\phi$ decay is governed by the weak 
decay of the heavy b-quark, $b \rightarrow d s \overline{s}$. The light 
antiquark of the $B$ meson is the spectator in the decay, being only slightly 
accelerated by the exchange of a hard gluon to form a pion in the final state.

In this section we present the basic ingredients for a
calculation of such a penguin-induced decay.
The first ingredient is the NLO effective weak Hamiltonian, which allows a 
consistent study of nonleptonic decays in which penguin operators are 
involved. The second ingredient, on which the paper is based, is a 
modified perturbative method for calculating exclusive decays by which the 
matrix elements of four-quark weak-Hamiltonian operators are perturbatively 
calculable.

\widetext

\subsection{Low-energy effective weak Hamiltonian beyond the leading 
logarithmic approximation}

Following Ref. \cite{BuJaLW} we consider the NLO effective weak Hamiltonian 
for $b \rightarrow d$ transitions:
\Beq
H_{\rm eff}(\Delta B = -1) = \frac{G_F}{\sqrt{2}} \sum_{q = u,c} \, 
V_q \, \left ( c_1(\mu) {\cal O}_1^{(q)} + c_2(\mu) {\cal O}_2^{(q)} + 
\sum_{k=3}^{10} c_k(\mu) {\cal O}_k \right ) \; .
\label{eq:ewH1}
\Eeq

The scale-dependent Wilson coefficients $c_i(\mu)$ are the short-distance 
part of the Hamiltonian and 
include NLO QCD corrections and leading-order 
$\alpha_{\rm em}$ corrections. With $V_q$ we denote products of CKM 
mixing matrix elements relevant to $b \rightarrow d$ transitions, 
$V_q = V_{qd}^{\ast} V_{qb}$. Local four-quark operators, renormalized at 
the scale $\mu$, are 
\Beqa
{\cal O}_1^{(q)} &=& (\overline{d}_{\alpha} q_{\beta})_{V-A} 
(\overline{q}_{\beta} b_{\alpha})_{V-A} \; , \;\;\;\;\;\;\;\;\;\;\;\;\;\;\;\; 
{\cal O}_2^{(q)}  =  (\overline{d}q)_{V-A} 
(\overline{q} b)_{V-A} \; , \nonumber \\
{\cal O}_3 &=& (\overline{d} b)_{V-A} \sum_{q'} 
(\overline{q'} q')_{V-A} \; , \;\;\;\;\;\;\;\;\;\;\;\;\;\;\;\;\;\; 
{\cal O}_4 =  (\overline{d}_{\alpha} b_{\beta})_{V-A} \sum_{q'} 
(\overline{q'}_{\beta} {q'}_{\alpha})_{V-A} \; , \nonumber \\
{\cal O}_5 &=& (\overline{d} b)_{V-A} \sum_{q'} 
(\overline{q'} q')_{V+A} \; , \;\;\;\;\;\;\;\;\;\;\;\;\;\;\;\;\;\; 
{\cal O}_6 =  (\overline{d}_{\alpha} b_{\beta})_{V-A} \sum_{q'} 
(\overline{q'}_{\beta} {q'}_{\alpha})_{V+A} \; , \nonumber \\
{\cal O}_7 &=& \frac{3}{2} (\overline{d} b)_{V-A} \sum_{q'} 
e_{q'} (\overline{q'} q')_{V+A} \; , \;\;\;\;\;\;\;\;\;\;\; 
{\cal O}_8 =  \frac{3}{2} (\overline{d}_{\alpha} b_{\beta})_{V-A} \sum_{q'} 
e_{q'}(\overline{q'}_{\beta} {q'}_{\alpha})_{V+A} \; ,  \nonumber \\
{\cal O}_9 &=& \frac{3}{2} (\overline{d} b)_{V-A} \sum_{q'} 
e_{q'} (\overline{q'} q')_{V-A} \; , \;\;\;\;\;\;\;\;\;\;  
{\cal O}_{10} =  \frac{3}{2} (\overline{d}_{\alpha} b_{\beta})_{V-A} \sum_{q'} 
e_{q'}(\overline{q'}_{\beta} {q'}_{\alpha})_{V-A}  \; ,
\label{eq:op}
\Eeqa
where $V\pm A = 1/2 \; \gamma_{\mu} (1\pm \gamma_5)$, $\alpha$ and $\beta$ are 
color indices, $q' \in \{u,d,s,c,b \}$, and $e_{q'}$ are the corresponding 
quark charges. ${\cal O}_1^{(q)}$ and ${\cal O}_2^{(q)}$ are tree-level 
operators, ${\cal O}_3 ,...,{\cal O}_6$ are QCD-penguin operators, 
and ${\cal O}_7 ,...,{\cal O}_{10}$ are EW penguins.

From the quark content of the operators 
${\cal O}_1^{(q)}$ and ${\cal O}_2^{(q)}$
it is obvious that they do not contribute at the tree level in 
$b \rightarrow d s \overline{s}$ transitions. 
Such transitions are pure penguin-induced, 
receiving contributions from the operators ${\cal O}_3 ,...,{\cal O}_{10}$, in 
which $q'$ is restricted to be a strange quark, $q' = s$.

\begin{figure}
\centerline{\epsfig{file=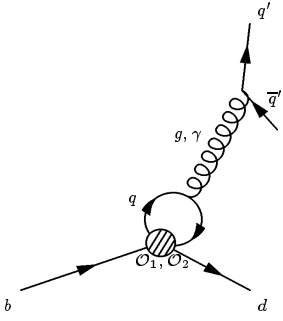,height=5.5cm,width=6.5cm,silent=}}
\caption{QCD and QED one-loop penguinlike contributions of the
tree-level operators ${\cal O}_1$ and ${\cal O}_2$ in the
$b \rightarrow d q' {\overline q'}$ decay.}
\label{f:genP}
\end{figure}

In the NLO weak Hamiltonian (\ref{eq:ewH1}) the renormalization-scheme 
dependence 
of the Wilson coefficients is explicitly canceled by the inclusion of the 
one-loop QCD and QED matrix elements of the tree-level operators 
${\cal O}_{1,2}^{(q)}$, Fig. \ref{f:genP}. The one-loop matrix elements of 
the NLO weak Hamiltonian for a $b \rightarrow d s \overline{s}$ transition 
can be written in terms of products of the tree-level matrix elements of 
penguin operators and the renormalization scheme-independent coefficients 
$\overline{c}(\mu)$ :

\Beqa
\langle d s \overline{s} &|& {\cal H}_{eff} (\Delta B = -1) | b \rangle =
\frac{G_F}{\sqrt{2}} \sum_{q = u,c} V_q \left ( \sum_{k=3}^{10} 
\overline{c}_k(\mu) \langle {\cal O}_k \rangle^{\rm tree} \right . \nonumber \\
& & -\frac{\alpha_s (\mu)}{24 \pi} 
\overline{c}_2(\mu) \left \{ \left \langle \left ( \frac{10}{9} - \Delta
G(m_q^2, q^2, \mu^2) \right )
\left ( {\cal O}_3  - 3 {\cal O}_4 +  {\cal O}_5 -3 {\cal O}_6 \right )
\right \rangle^{\rm tree} \right \} \nonumber \\
& & \left . + \frac{\alpha_{\rm em}}{9 \pi} 
( 3 \overline{c}_1(\mu) + \overline{c}_2(\mu)) 
\left \{ \left \langle  \left ( \frac{10}{9} -       
\Delta G(m_q^2, q^2, \mu^2) \right )
( {\cal O}_7 + {\cal O}_9 ) 
\right \rangle^{\rm tree} \right \} \right ) \; ,
\label{eq:MEfull}
\Eeqa
where $\langle {\cal O}_k \rangle^{\rm tree} = \langle d s \overline{s} |
{\cal O}_k | b \rangle^{\rm tree}$ .

The function $\Delta G (m_q^2, q^2,\mu^2)$ arises from the one-loop penguinlike 
diagrams with $q = u,c$ quarks in the loop:
\Beq
\Delta G(m_q^2, q^2,\mu^2) = -4\int_0^1 du\,u (1-u) {\rm ln} \left 
( \frac{m_q^2 - q^2 u (1-u)}
{\mu^2} \right)  \; .
\label{eq:dG}
\Eeq
Here, as well as the quark mass $m_q$ and the renormalization scale $\mu$, 
there appears a new parameter $q^2$, which is the momentum squared 
of a virtual gluon or photon 
emerging from the loop. In the $b \rightarrow d s \overline{s}$ transition 
$q^2$ can be indentified with the sum of the strange quark momenta squared. 

Concentrating now on the specific process $B^{-}(b \overline{u}) \rightarrow 
\pi^{-}(d \overline{u}) \phi (s \overline{s})$, a few comments are in order. 
The strange quark and antiquark building $\phi$-meson are obviously in the 
color-singlet state. The $s \overline{s}$ pair coming from the virtual gluon 
decay in the QCD-penguin diagrams builds a color-octet state. Therefore, first, 
one expects a small contribution from the QCD-penguin operators and, 
second, the one-loop QCD penguinlike contribution (shown in Fig. \ref{f:genP} 
with 
an exchanged gluon) is not present in the $B^{-}\rightarrow \pi^{-} \phi$ decay. 
The final expression for the matrix element is then
\Beqa
\langle \pi^{-} \phi &|& {\cal H}_{eff} (\Delta B = -1) | B^{-} \rangle =
\frac{G_F}{\sqrt{2}} \sum_{q = u,c} V_q \left ( \sum_{k=3}^{10}
\overline{c}_k(\mu) \langle {\cal O}_k \rangle^{\rm tree} \right .\nonumber \\
& & \left . + \frac{\alpha_{\rm em}}{9 \pi} \left ( \frac{10}{9} -
\Delta G(m_q, \langle q^2 \rangle, \mu) \right ) ( 3 \overline{c}_1(\mu) + 
\overline{c}_2(\mu))
\left \{\langle {\cal O}_7 \rangle^{\rm tree} + \langle {\cal O}_9
\rangle^{\rm tree} \right \} \right ) \; .
\label{eq:ME}
\Eeqa
Here $\langle {\cal O}_k\rangle^{\rm tree} \simeq \langle \pi^{-}\phi |
{\cal O}_k | B^{-} \rangle$. 

\narrowtext

We have retained the $\alpha_{\rm em}$-propotional 
term as a part of the NLO weak Hamiltonian, although we show later that, 
in the perturbative approach to the 
order we are working with, such a 
contribution emerges naturally. This term is producing hard final-state 
interaction phase shifts, necessary for generating CP asymmetry, which are 
due to the 
on-shell quarks rescattering in the loop for particular values of $q^2$. 
The CP asymmetry depends strongly on the value of $q^2$ and, 
in the factorization approach, the major 
source of uncertainties in predictions comes from 
the lack of information about the $q^2$ value after the factorization of 
hadronic matrix elements is performed. On the contrary, 
in the perturbative approach, the $q^2$-dependence 
of the loop amplitude is calculable directly 
as a part of the hadronic matrix element and it is determined by 
the momentum distributions in a particular process. 
However, in the strict factorization, and in the perturbative calculation 
for the process considered, the average $q^2$-value can be simply determined 
to be the mass of the $\phi$ meson squared, 
$\langle q^2 \rangle = M^2_{\phi}$. Further discussion about this point 
is left for Sec. V.

Let us now continue with the estimation of the matrix elements of four-quark 
operators.

\subsection{Modified perturbative approach to the calculation of the 
matrix elements of four-quark operators}

Perturbative calculations of matrix elements in exclusive hadron decays 
can be carried out in the Brodsky-Lapage (BL) formalism \cite{BLetc}. 
Hadrons are considered in 
the leading approximation as a bound state of valence quarks and/or antiquarks, 
depending on the hadron. The amplitude of the process factorizes into the 
convolution of distribution amplitudes of hadrons involved in the decay 
(hadron wave functions) and the hard scattering amplitude of valence partons. 
Hadronic wave functions represent the nonperturbative part, which has to be 
determined for heavy hadrons in relativistic constituent or nonrelativistic 
models, or for light hadrons by the QCD sum-rule method or by lattice 
calculations. Further discussion about the wave functions is given in 
Sec. IV.

\begin{figure}
 \centerline{\epsfig{file=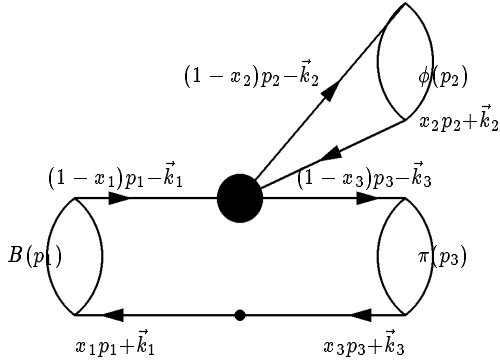,height=6.5cm,width=7.5cm,silent=}}
\caption{The basic graph of the $B^{-} \rightarrow \pi^{-} \phi$ decay with the
momentum definitions specified. The black circle stands for the NLO effective
weak Hamiltonian (\protect\ref{eq:ewH1}).}
\label{f:basic}
\end{figure}

The hard scattering amplitude can be calculated perturbatively, taking into 
account all possible exchanges of a hard gluon between valence partons in a 
given $\alpha_s$-order of the calculation. Valence quarks are carrying some 
fraction of momentum of their parent hadron and the final expression is 
integrated over the fractions $x_i$ ($i=1,2,3$), see Fig. \ref{f:basic}. 
The concern about the applicability of this 
method to exclusive processes was raised when it was noted that even 
at large momentum transfer the contribution to these processes could come 
predominantly from the momentum regions in which $\alpha_s$ is large
\cite{Ra}, \cite{IsLl}. The problem is connected with the end-point region 
of momentum 
fractions. Namely, when one of the hadron constituents carries all the 
momentum of a hadron, the situation is no more perturbative and significant 
uncontrollable soft contributions might appear. The solution of the 
problem was proposed by Li and Sterman \cite{LiSt}. 
Contrary to the BL formalism, they suggested to go beyond the collinear approximation 
and to retain the small transverse momentum of valence quarks. 
Owing to the transverse degrees of freedom, the parton virtualities
become large enough in the whole region for a reliable perturbative 
calculation. 
Furthermore, they 
included the Sudakov form factor for each of the hadrons in the decay to 
suppress the contributions from dangerous soft regions. All these effects 
can be incorporated into the factorization formula by expressing the transverse 
momentum variables in the Fourier transformed $b$-space. 

\widetext

The very last formula,  
which is used throughout this paper to calculate the matrix elements 
of four-quark operators from (\ref{eq:ME}) relevant to the 
$B^{-}\rightarrow \pi^{-} \phi$ decay, is 
\Beq
\langle \pi^{-} \phi | {\cal O}_{k}| B^{-} \rangle = \int [dx] \int
\left [ \frac{d^2\vec{b}}{4 \pi} \right ] \Psi_{\pi}^{\ast}(x_3, \vec{b}_3) 
\Psi_{\phi}^{\ast}(x_2, \vec{b}_2) \, T_k (\{ x \}, \{ \vec{b} \}, 
M_B ) 
\Psi_{B}(x_1, \vec{b}_1) \displaystyle{e^{-
\displaystyle{S \left ( \{ x \}, \{ \vec{b} \}, M_B \right )} }} \; ,
\label{eq:MEop}
\Eeq
where $x_i$ $(i = 1,2,3)$ are fractions of longitudinal momenta of B, $\phi$,  
and $\pi$ mesons, respectively. Analogously, $b_i$ denote the Fourier-
transformed transverse momenta of these mesons. $[dx] = dx_1\,dx_2\,dx_3$ 
 and $\{x\}$ denotes the set of variables $\{x_1,\, x_2, \,x_3\}$. 
Similarly, for $\vec{b}$ variables. 
$\Psi_{\pi}^{\ast}$, $\Psi_{\phi}^{\ast}$, and 
$\Psi_{B}$ are the wave functions of the outgoing $\pi$ and $\phi$ mesons, and 
the decaying $B$ meson, respectively. $T_k$ is the hard scattering amplitude 
describing the $B \rightarrow \pi \phi$ decay at the one-loop level caused by one of 
the operators ${\cal O}_k$ (\ref{eq:op}). 
The exponential factor in formula (\ref{eq:MEop}) is the Sudakov factor. 
Its explicit form is given in Sec. IV.

\narrowtext

\section{Calculation of the $B^{-} \rightarrow \pi^{-} \phi$ decay}

The basic graph representing the $B^{-} \rightarrow \pi^{-} \phi$ decay in the 
perturbative approach is shown in Fig. \ref{f:basic}.  

In order to perform the calculation, we have to specify the hadronic momenta. 
The simplest choice can be made using the light-cone coordinates and taking 
the $B$ meson to be at rest:
\Beqa
P_B &=& p_1 = \frac{M_B}{\displaystyle \sqrt{2}}\,(1,1,\vec{0})\; ,
\;\;\;\;\; p_1^2 = M_B^2\; ,  \nonumber \\
P_{\phi} &=& p_2 = \frac{M_B}{\displaystyle \sqrt{2}}\,(1,r^2,\vec{0})\; ,
\;\;\;\;\; p_2^2 = M_{\phi}^2 \; , \nonumber \\
P_{\pi} &=& p_3 = \frac{M_B}{\displaystyle \sqrt{2}}\,(0,1-r^2,\vec{0})\; , 
\;\;\;\;\; p_3^2 = 0 \; .
\label{eq:kin}
\Eeqa
Here, $M_B$ and $M_{\phi}$ are the masses of the $B$ and the $\phi$ meson, respectively,  
and $r$ is defined 
as their ratio, $r = M_{\phi}/M_B$. The pion is taken to be massless. 
In addition to carrying some portion of the momentum of 
their parent meson, the valence quarks also carry some small 
transverse momentum $\vec{k}$ (see Fig. \ref{f:basic}).

\begin{figure}
 \centerline{\epsfig{file=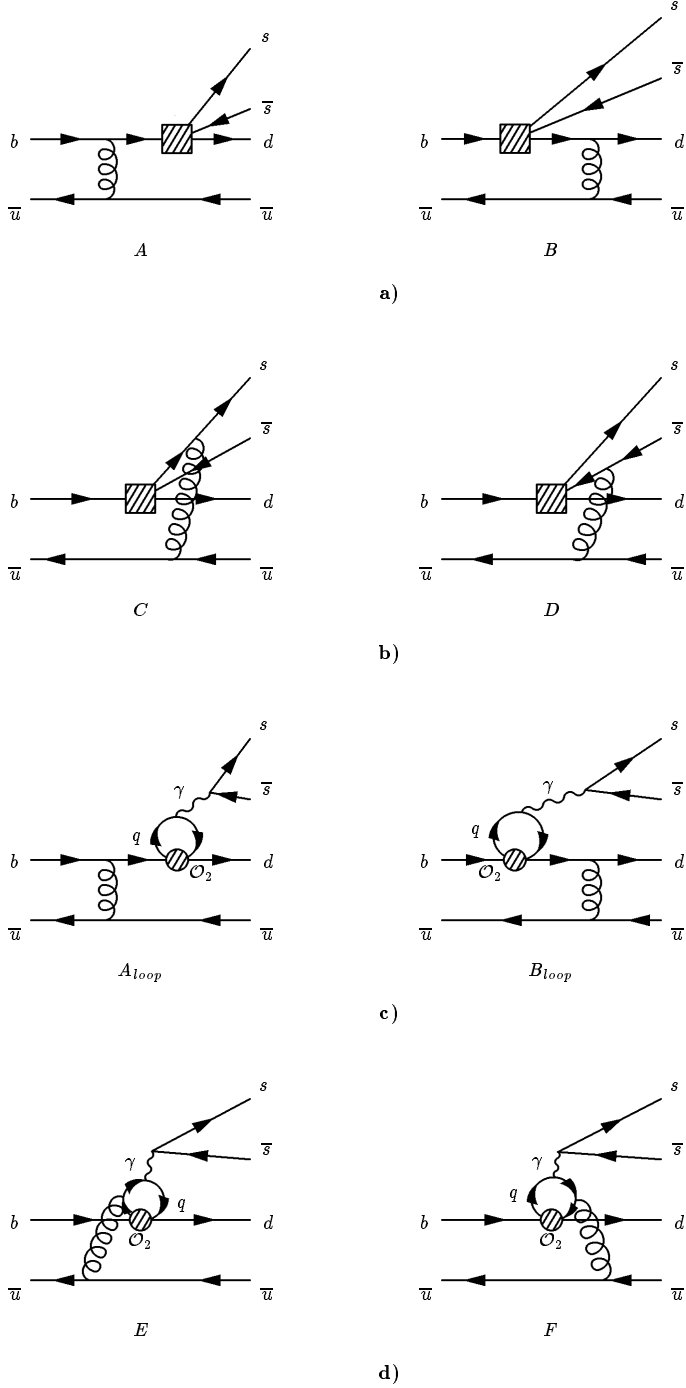,height=22.5cm,width=16.5cm,silent=}}
\caption{Leading-order contributions to the $B^{-} \rightarrow \pi^{-} \phi$
decay:
{\bf a)} factorizable, {\bf b)} nonfactorizable, and {\bf c)} and {\bf d)}
penguinlike.
The square stands for the penguin operators ${\cal O}_k$, $k=3,...,10$, and
the circle represents the tree-level operator ${\cal O}_2$.}
\label{f:main}
\end{figure}

Working in the leading order, the perturbative part, namely, the amplitudes of 
the four-quark operators (\ref{eq:MEop}), can be calculated from the Feynman graphs with 
all possible attachments of a hard gluon, shown in Fig. \ref{f:main}. As 
discussed later explicitly, Fig. \ref{f:main}(a) shows factorizable 
amplitudes of the decay. Figure \ref{f:main}(b) shows nonfactorizable 
amplitudes, 
not presented in the BSW-factorization approach. Contributions of one-loop 
induced penguinlike diagrams coming from the tree-level operators are 
shown in Fig. \ref{f:main}(c). These contributions can be taken into account 
immediately assuming the NLO weak Hamiltonian, but they emerge 
naturally in the pertubative model. Finally, Fig. \ref{f:main}(d) shows 
some additional diagrams that have to be included to perform a proper 
${\cal O}(\alpha_s \alpha_{\rm em})$ calculation.

In full hadron wave functions one can split up spin wave functions from 
the rest; particularly,
\Beqa
\Psi_B(x_1, \vec{b}_1) &=& \frac{1}{\sqrt{2}} ({\not p_1} + M_B)\gamma_5 
\frac{\openone_c}{\sqrt{3}} \Phi_B(x_1, \vec{b}_1) \; ,\nonumber \\
\Psi_{\phi}^{\ast}(x_2, \vec{b}_2) &=& \frac{1}{\sqrt{2}}{\not \epsilon} 
({\not p_2} + M_{\phi})
\frac{\openone_c}{\sqrt{3}} \Phi_{\phi}^{\ast}(x_2, \vec{b}_2)\; , \nonumber \\
\Psi_{\pi}^{\ast}(x_3, \vec{b}_3) &=& \frac{1}{\sqrt{2}} \gamma_5 {\not p_3}
\frac{\openone_c}{\sqrt{3}} \Phi_{\pi}^{\ast}(x_3, \vec{b}_3) \; ,
\label{eq:WFfull}
\Eeqa
where $\openone_c$ is the unit color matrix, $\epsilon^{\mu}$ is the polarization 
vector of the $\phi$ meson. The scalar wave functions 
$\Phi$ are specific to each of the mesons and are discussed in Sec. IV.

In this process it is enough to calculate QCD-penguin contributions, because 
EW-penguin operators are simply related to QCD penguins as
\Beqa
{\cal O}_3 &=& (\overline{d} b)_{V-A} 
(\overline{s} s)_{V-A} = \frac{2}{3 e_s} {\cal O}_9 \; , \nonumber \\
{\cal O}_4 &=& - (\overline{d} s)_{V-A}
(\overline{s} b)_{V-A} = \frac{2}{3 e_s} {\cal O}_{10} \; ,
\nonumber \\
{\cal O}_5 &=& (\overline{d} b)_{V-A} 
(\overline{s} s)_{V+A} = \frac{2}{3 e_s} {\cal O}_7 \; , \nonumber \\
{\cal O}_6 &=& -2 (\overline{d} s)_{S+P} 
(\overline{s} {b})_{S-P}= \frac{2}{3 e_s} {\cal O}_8 \; , 
\label{eq:opQE}
\Eeqa
where $e_s$ is the strange quark charge, $e_s = - 1/3 e$.

Applying the standard procedure for calculating Feynman diagrams, the 
hard scattering amplitudes $T_k$ can be easily worked out by performing color 
and spin traces. They appear after the spin and color parts of the wave 
functions 
are added to the amplitude, leaving the scalar functions $\Phi$ at the place 
of the full mesonic wave functions $\Psi$ in (\ref{eq:MEop}).

\widetext

All operators receive the same contribution from the factorizable 
diagrams, Fig. \ref{f:main}(a):
\Beqa
T_{fact} &=& T_A + T_B \; , 
\nonumber \\
T_A &=& -4\pi \alpha_s(\mu) C_f \frac{N_c}{\sqrt{3}} \frac{1}{2\sqrt{2}} 
32 M_B^3 r \left [ (1 + x_3 (1-r^2)) \epsilon\cdot p_3 \right ]
\cdot \frac{1}{D_G} \frac{1}{D_b} \; ,
\nonumber \\
T_B &=& -4\pi \alpha_s(\mu) C_f \frac{N_c}{\sqrt{3}} \frac{1}{2\sqrt{2}} 
32 M_B^3 r \left [ x_1 (1-r^2) \epsilon\cdot p_1 -
x_1 \epsilon\cdot p_3 \right ] \cdot
\frac{1}{D_G} \frac{1}{D_{q_1}} \; , 
\label{eq:fact}
\Eeqa
where $C_f$ is the color factor equal to $C_f = C_F = 4/3$ for the ${\cal O}_3$ 
(${\cal O}_9$) and ${\cal O}_5$ (${\cal O}_7$) operators and $C_f = C_F/N_c$ 
for the others. We are going to take the number of colors $N_c$ to be equal 
to 3.

$D_G$, $D_b$, and $D_{q_1}$ denote the denominators of the exchanged gluon, and 
of the virtual $b$ and $d$ quark propagators, respectively:
\Beqa
D_G &=& q_G^2 + i\eta = M_B^2 [x_1^2 - x_1 x_3 (1-r^2)] - (\vec{k}_1 - 
\vec{k}_3)^2 + i \eta \; , \nonumber \\
D_b &=& q_b^2 + i\eta \simeq  
-M_B^2 x_3 (1-r^2) - {\vec{k}_3}^2 + i\eta \; , \nonumber \\
D_{q_1} &=& q_{q_1}^2 + i\eta = M_B^2 [x_1^2 - x_1 (1-r^2)] - 
{\vec{k}_1}^2 +i\eta \; .
\label{eq:prop1}
\Eeqa
The approximation made here was to take $m_b \simeq M_B$ in the propagator 
of the heavy $b$ quark. Other propagators are massless.

The operators ${\cal O}_4$ (${\cal O}_{10}$) and ${\cal O}_6$ (${\cal O}_8$) 
also receive 
nonfactorizable contributions coming from the estimation of the diagrams 
in Fig. \ref{f:main}(b). The expressions for the ${\cal O}_4$ (${\cal O}_{10}$) 
operator are
\Beqa
T_{nonfact} ({\cal O}_4) &=& T_C({\cal O}_4) + T_D({\cal O}_4) \; , 
\nonumber \\
T_C({\cal O}_4) &=& -4\pi \alpha_s(\mu) \frac{C_F}{3} \frac{N_c}
{\sqrt{3}} \frac{1}{2\sqrt{2}} 32 M_B^3 r
\left [ (1-r^2)(1-x_1-x_2) \epsilon \cdot p_1 \right ] \cdot 
\frac{1}{D_G} \frac{1}{D_{q_2}} \; ,
\nonumber \\
T_D({\cal O}_4) &=& -4\pi \alpha_s(\mu) \frac{C_F}{3} 
\frac{N_c}{\sqrt{3}} \frac{1}{2\sqrt{2}} 32 M_B^3 r
\left [ (2 x_1 -x_2-x_3-r^2 (x_2-x_3)) \epsilon \cdot p_3 \right ] \cdot 
\frac{1}{D_G} \frac{1}{D_{q_3}} \; , 
\label{eq:nf4}
\Eeqa
and similarly for the ${\cal O}_6$ (${\cal O}_8$) operator
\Beqa
T_{nonfact} ({\cal O}_6) &=& T_C({\cal O}_6) + T_D({\cal O}_6) \; ,  
\nonumber \\
T_C({\cal O}_6) &=& -4\pi \alpha_s(\mu) \frac{C_F}{3} \frac{N_c}
{\sqrt{3}}
\frac{1}{2\sqrt{2}} 32 M_B^3 r  \nonumber \\
& & \times \left [ (1-2 x_1 - x_2 +x_3 + r^2 (1-x_2-x_3))\epsilon \cdot p_3 
\right ]\cdot \frac{1}{D_G} \frac{1}{D_{q_2}} \; , 
\nonumber \\
T_D({\cal O}_6) &=& -4\pi \alpha_s(\mu)  \frac{C_F}{3} 
\frac{N_c}{\sqrt{3}} \frac{1}{2\sqrt{2}} 32 M_B^3 r
\left [ (1-r^2) (x_1-x_2)\epsilon \cdot p_1 \right ]\cdot
\frac{1}{D_G} \frac{1}{D_{q_3}} \; . 
\label{eq:nf6}
\Eeqa

The denominators of the virtual quark propagators in Fig. \ref{f:main}(b) are
\Beqa
D_{q_2} &=& q_{q_2}^2 + i\eta = M_B^2 \left [ (1-x_1-x_3)(-(x_1-x_3)+ 
(1-x_2-x_3) r^2) \right ] -
(\vec{k}_1 + \vec{k}_2 - \vec{k}_3)^2 + i\eta \; , \nonumber \\
D_{q_3} &=& q_{q_3}^2 + i\eta = M_B^2 \left [ (x_1-x_2)((x_1-x_3)-
(x_2-x_3) r^2) \right ] -
(\vec{k}_1 - \vec{k}_2 - \vec{k}_3)^2 + i\eta  \; . 
\label{eq:prop2}
\Eeqa

The calculation of the one-loop EW-penguinlike contributions from 
Fig. \ref{f:main}(c) follows the already familiar procedure. Both diagrams 
$A_{loop}$ and $B_{loop}$  are propotional to their skeleton graphs $A$ and $B$ 
, respectively. Performing renormalization consistent with the use of the 
NLO weak 
Hamiltonian and its renormalization-scheme independence \cite{Fl1}, 
the result, as expected, is given by 
\Beq
-3 T_{fact} \cdot \frac{\alpha_s}{9 \pi} \left 
( \frac{10}{9} - \Delta G(m_q^2,M^2_{\phi},\mu^2) \right ) \; , 
\label{eq:factdG}
\Eeq
where $\Delta G$ has already been defined by (\ref{eq:dG}) and the value of 
the parameter $q^2$ is 
determined from momentum distributions in the process to be $M_{\phi}^2$. 

The contributions from other two penguinlike diagrams, Fig. \ref{f:main}(d), 
are 
lengthy because they involve the $b \rightarrow d \gamma^{\ast} g^{\ast}$ 
vertex calculation \cite{SiWy} and will be given only in the final 
form, in expression (\ref{eq:aTloop}), and in the Appendix.

The next step to be performed is to express 
the hard scattering amplitudes,
 (\ref{eq:fact},\ref{eq:nf4},\ref{eq:nf6}) in the Fourier-transformed space 
of transverse momenta. The Fourier-transformed amplitudes read
\Beqa
{\tilde T}_{fact} &=& - \frac{\alpha_s(\mu)}{\pi}  \cdot {C_f} \frac{N_c}
{\sqrt{3}} \frac{1}{2\sqrt{2}}
32 M_B^3 r \left \{ \left [(1+x_3 (1-r^2)) 
\epsilon\cdot p_3 \right ]\, h_A (D_G,D_b, b_1,b_3) \right .\nonumber \\
&+& 
\left .
\left [ x_1(1-r^2) \epsilon\cdot p_1 - x_1 
\epsilon\cdot p_3) \right ] \, h_B (D_G,D_{q_1}, b_3,b_1)\right \} \; , 
\nonumber \\
{\tilde T}_{nonfact}({\cal O}_4) &=& - \frac{\alpha_s(\mu)}{\pi}  \cdot \frac{C_F}{3} 
\frac{N_c} {\sqrt{3}} \frac{1}{2\sqrt{2}} 32 M_B^3 r \left \{ 
\left [ (1-r^2)(1-x_1-x_2) \epsilon \cdot p_1 \right ]\,h_C(D_{q_2},D_G,b_2,b_1)
\right .  \nonumber \\
&+&
\left .
\left [ (2 x_1 -x_2-x_3-r^2 (x_2-x_3)) \epsilon \cdot p_3 \right ]\,
h_D (D_{q_3},D_G,b_2,b_1) \right \} \; , 
\nonumber \\
{\tilde T}_{nonfact}({\cal O}_6) &=& -\frac{\alpha_s(\mu)}{\pi}  \cdot \frac{C_F}{3} 
\frac{N_c} {\sqrt{3}} \frac{1}{2\sqrt{2}} 32 M_B^3 r \left \{
\left [ (1-2 x_1 - x_2 +x_3 + r^2 (1-x_2-x_3))\epsilon \cdot p_3 \right ]\,
\right . \nonumber \\
& &  \cdot \left . h_C(D_{q_2},D_G,b_2,b_1)
+
\left [ (1-r^2) (x_1-x_2)\epsilon \cdot p_1 \right ]\,
h_D (D_{q_3},D_G,b_2,b_1) \right \} \; , 
\label{eq:Tall} 
\Eeqa
where
\Beqa
h_A (D_G, D_{b}, b_1,b_3) &=& K_0 \left (\sqrt{-\displaystyle q_G^2} 
|\vec{b}_1|\right ) 
K_0 \left (\sqrt{-q_b^2}|\vec{b}_1 + \vec{b}_3|\right ) \delta(\vec{b}_2)\; , 
\nonumber \\
h_B (D_G, D_{q1}, b_3,b_1) &=& K_0 \left (\sqrt{-\displaystyle q_G^2} 
|\vec{b}_3|\right ) 
K_0 \left (\sqrt{-q_{q1}^2} |\vec{b}_1 + \vec{b}_3| \right ) 
\delta(\vec{b}_2) \; ,  
\nonumber \\
h_C (D_{q_2},D_G,b_2,b_1) &=& K_0 \left (\sqrt{-\displaystyle q_{q_2}^2} 
|\vec{b}_2|\right ) 
K_0 \left (\sqrt{-q_{G}^2} |\vec{b}_1 - \vec{b}_2|\right ) 
\delta(\vec{b}_1 +\vec{b}_3) \; , \nonumber \\ 
h_D (D_{q_3},D_G,b_2,b_1) &=& K_0 \left (\sqrt{-\displaystyle q_{q_3}^2} 
|\vec{b}_2|\right ) 
K_0 \left (\sqrt{-q_{G}^2} |\vec{b}_1 - \vec{b}_2|\right ) 
\delta(\vec{b}_1 +\vec{b}_3) \; , 
\Eeqa
and $K_0$ is the modified Bessel function of order zero.

For the scales $\mu$ appearing in formulas (\ref{eq:Tall}) and 
(\ref{eq:factdG}) we are going to take the largest mass scale in 
the particular diagram :
\Beqa
t_A &=& max \left (\sqrt{-q_G^2}, \sqrt{-q_b^2}, 1/b_1, 1/b_3
\right ) \; , \quad
t_B = max \left (\sqrt{-q_G^2}, \sqrt{-q_{q_1}^2}, 1/b_1, 1/b_3
\right ) \; , \nonumber \\
t_C &=& max \left (\sqrt{-q_G^2}, \sqrt{-q_{q_2}^2}, 1/b_1, 1/b_2 
\right ) \; , \quad
t_D = max \left (\sqrt{-q_G^2}, \sqrt{-q_{q_3}^2}, 1/b_1, 1/b_2 
\right ) \; , \nonumber \\
t_{loop} &=& max \left (\sqrt{-q_G^2}, 1/b_1 \right ) \, .
\label{eq:texp}
\Eeqa
which ensures the reliable perturbative calculations with the small 
$\alpha_s$-coupling.

Performing trivial $b$ integrals over $\delta$ functions and performing 
angular integrations by using Graph's theorem:
\Beqa
f(x,b_1,b_2) &=& \int d\phi K_0 (x|\vec{b}_1 \pm \vec{b}_2|) \nonumber \\
&=& 2\pi [\Theta (b_1-b_2) K_0 (x b_1) I_0 (x b_2) + 
\Theta (b_2-b_1) K_0 (x b_2) I_0 (x b_1) ] \; , 
\label{eq:fexp}
\Eeqa
one can finally write
the total amplitude of the $B^{-} \rightarrow \pi^{-} \phi$ decay as
\Beqa
{\cal M} &=& \langle \pi \phi | H_{eff} | B \rangle = \frac{G_F}{\sqrt{2}}
\sum_{q=u,c} V_q {\cal A}_q  \; , \nonumber \\
\nonumber \\
{\cal A}_q &=& \left \{ 3 (\overline{c}_3 + \overline{c}_5 - \frac{1}{2}
(\overline{c}_7 + \overline{c}_9) ) + \overline{c}_4 + \overline{c}_6 
 - \frac{1}{2} (\overline{c}_8 + \overline{c}_{10})\right \} 
\langle T_{fact} \rangle  \nonumber \\
&+& (\overline{c}_4 - \frac{1}{2}\overline{c}_{10}) \langle 
T_{nonfact}({\cal O}_4) \rangle + (\overline{c}_6 - \frac{1}{2}
\overline{c}_{8}) \langle 
T_{nonfact}({\cal O}_6) \rangle  \nonumber \\
&-& \frac{\alpha_{\rm em}}{9 \pi}(3 \overline{c}_1 + \overline{c}_2)\cdot 3 \langle
T_{fact} \rangle \left ( \frac{10}{9} - \Delta 
G(m_q^2, M_{\phi}^2, m_b^2) \right ) \nonumber \\
&-& \frac{2 \alpha_{\rm em}}{3 \pi}\, \overline{c}_2 \langle 
T_{loop}\rangle_q \; , 
\label{eq:Mamp}
\Eeqa
with the matrix elements
\Beqa
\langle T_{fact} \rangle &=& -\frac{C_F}{3} f_{\phi} M_B^3 r \int dx_1 dx_3 \,
\int b_1 db_1\, b_3 db_3 \Phi_{B}(x_1,b_1) \Phi_{\pi}^{\ast}(x_3,b_3)\, 
\nonumber \\
&\times& \left \{
\frac{\alpha_s(t_A)}{\pi} H(D_G,D_b,b_1,b_3)\,
\left [(1+x_3 (1-r^2)) \epsilon\cdot p_3 \right ]\, 
e^{- \displaystyle (S_B(t_A) + 
S_{\pi}(t_A) )} \right . \nonumber \\
& & \left .  +\frac{\alpha_s(t_B)}{\pi} H(D_G,D_{q_1},b_3,b_1)\,
\left [ x_1(1-r^2) \epsilon\cdot p_1 - x_1
\epsilon\cdot p_3 \right ] \, e^{- \displaystyle (S_B(t_B) + 
S_{\pi}(t_B) )}    \right \} \; ,  
\nonumber \\
\langle T_{nonfact}({\cal O}_4) \rangle &=& 
-\frac{C_F}{3} f_{\phi} M_B^3 r \int
[dx]\, \int b_1 db_1\, b_2 db_2 \Phi_{B}(x_1,b_1) \frac{\displaystyle 
{\tilde \Phi}_{\phi}^{\ast}(x_2,b_2)}{4 \pi} \Phi_{\pi}^{\ast}(x_3,b_1)\,
 \nonumber \\
&\times& \left \{
\frac{\alpha_s(t_C)}{\pi} H(D_{q_2},D_b,b_2,b_1)\,
\left [ (1-r^2)(1-x_1-x_2) \epsilon \cdot p_1 \right ] 
\; e^{- \displaystyle (S_B(t_C) + S_{\phi}(t_C) +
S_{\pi}(t_C) ) |_{b_3 = b_1}}
\right .  \nonumber \\
& & 
\left .
 + \frac{\alpha_s(t_D)}{\pi} H(D_{q_3},D_b,b_2,b_1)\,
\left [ (2 x_1 -x_2-x_3-r^2 (x_2-x_3)) \epsilon \cdot p_3 \right ] 
\; \right .  \nonumber \\
& & \left . 
\cdot \, e^{- \displaystyle (S_B(t_D) + S_{\phi}(t_D) +
S_{\pi}(t_D) ) |_{b_3 = b_1}}
\right \}  \; , 
\nonumber \\
\langle T_{nonfact}({\cal O}_6) \rangle &=& 
-\frac{C_F}{3} f_{\phi} M_B^3 r \int
[dx]\, \int b_1 db_1\, b_2 db_2 \Phi_{B}(x_1,b_1) \frac{\displaystyle
{\tilde \Phi}_{\phi}^{\ast}(x_2,b_2)}{4 \pi} \Phi_{\pi}^{\ast}(x_3,b_1)\,
\nonumber \\
&\times& \left \{
\frac{\alpha_s(t_C)}{\pi} H(D_{q_2},D_b,b_2,b_1)\,
\left [ (1-2 x_1 - x_2 +x_3 + r^2 (1-x_2-x_3))\epsilon \cdot p_3 \right ]
\right . \nonumber \\
& & \cdot \left .
\;e^{- \displaystyle (S_B(t_C) + S_{\phi}(t_C) +
S_{\pi}(t_C) ) |_{b_3 = b_1}}
\right . \nonumber \\
& & \left .
+ \frac{\alpha_s(t_D)}{\pi} H(D_{q_3},D_b,b_2,b_1)\,
\left [ (1-r^2) (x_1-x_2)\epsilon \cdot p_1 \right ] 
\;e^{- \displaystyle (S_B(t_D) + S_{\phi}(t_D) +
S_{\pi}(t_D) ) |_{b_3 = b_1}}
\right \} \; , 
\label{eq:aTall}
\Eeqa
where the Fourier-transformed expressions for the propagators in Eqs. 
({\ref{eq:aTall}) have the general form
\Beq
H(D_1,D_2,b_1,b_2) = K_0(\sqrt{-D_1} b_1) f(\sqrt{-D_2}, b_1, b_2) \; ,
\label{eq:FTh}
\Eeq
with the function $f$ as defined by (\ref{eq:fexp}).

In (\ref{eq:Mamp}) we have also included the one-loop contributions from 
the diagrams of Figs. \ref{f:main}(c) and \ref{f:main}(d). 
The matrix elements receiving the contributions from the diagrams in 
Fig. \ref{f:main}(d) are 
\Beqa
\langle T_{loop} \rangle_q &=&   -\frac{C_F}{3} f_{\phi} M_B^3 r \int
dx_1 dx_3 \, \int b_1 db_1 \,\Phi_{B}(x_1,b_1) \Phi_{\pi}^{\ast}(x_3,b_1)\,
\nonumber \\
& & \times \frac{1}{M_{\phi}^2}\left \{
\frac{\alpha_s(t_{loop})}{\pi} K_0 \left ( \sqrt{-q_G^2} b_1 \right )
\left [ T^E_q + T^F_q \right ] 
e^{- \displaystyle (S_B(t_{loop}) + S_{\pi}(t_{loop}) ) |_{b_3 = b_1}}
\right \} \; .
\label{eq:aTloop}
\Eeqa
The expressions for $T^E_q$ and $T^F_q$ are given explicitly in the Appendix.

\narrowtext

One should note that in the above expressions we have pulled out the 
normalization factor $f_{\phi}/
{2 \displaystyle \sqrt{6}}$ ($f_{\phi}$ is the $\phi$-meson decay constant)
of the $\phi$ wave function and we denote the rest 
by $\tilde{\Phi}_{\phi}$ in order to have the same prefactor in both 
factorizable and nonfactorizable contributions. 

From the expression for $\langle T_{fact} \rangle$ in 
Eq.(\ref{eq:aTall}) it is easy to essentially recognize 
the factorization structure
in which a matrix element of a four-quark operator factorizes in the product of 
two current matrix elements 
$\langle \phi|(\overline{s} s)_{V-A}|0 \rangle \cdot \langle \pi^{-}(p_3)|
(\overline{d} b)_{V-A}|B^{-}(p_1) \rangle \sim f_{\phi} \epsilon^{\mu} 
\cdot (A p_1^{\mu} + B p_3^{\mu})$, and the ${\phi}$-meson wave function 
integrates out.  The current matrix element $\langle \pi^{-}(p_3)|
(\overline{d} b)_{V-A}|B^{-}(p_1) \rangle$ exactly describes the $B 
\rightarrow \pi$ transtion form factor at the momentum transfer $p^2 = 
(p_1 - p_3)^2 = M_{\phi}^2$. In Sec. IV we use this form factor 
to select mesonic wave functions. 

The expressions for the Sudakov exponents $S_{\pi}$, $S_{\phi}$, and $S_B$ 
in Eqs.(\ref{eq:aTall}) and (\ref{eq:aTloop}) are given in Sec. IV.

\section{Mesonic wave functions and Sudakov factors}

The calculation of the matrix elements requires the knowledge of the 
scalar meson wave functions $\Phi$. The hadronic wave functions represent the 
most speculative part of the perturbative approach. They are of nonperturbative 
origin and should be a universal, process-independent quantity. 

However, even for the most theoretically  and experimentally  exploited 
hadron, namely, the pion there are contradictory conclusions about the specific form 
of $\Phi_{\pi}$.
Theoretical calculations performed by using the QCD sum-rule method \cite{CZetc} and on the 
lattice \cite{GoKretc} cannot distinguish between the most promising forms of 
the pion wave function, the asymptotic one 
\Beq
\Phi_{\pi}^{as}(x) = 6 x(1-x) \frac{f_{\pi}}{2 \sqrt{6}}, 
\label{eq:Pias}
\Eeq
and the Chernyak-Zhitnitsky (CZ) wave function \cite{CZetc}
\Beq
\Phi_{\pi}^{CZ}(x) = 30 x(1-x) (1-2 x)^2 \frac{f_{\pi}}{2 \sqrt{6}}\; , 
\label{eq:Picz}
\Eeq
both normalized to satisfy the experimentally obtained value for the pion decay 
constant $f_{\pi} = 0.133 GeV$.
Besides, any comparison between theoretically calculable processes and 
existing experiments cannot provide an unambiguous determination between 
them \cite{HuMaS}.

\widetext

What we are going to regard as the CZ wave function throughout the paper 
is the CZ form in which the evolution from the hadronic
scale $\mu_{0} \sim 0.5 GeV$ to some scale $\mu_1$ is included \cite{BrLa}:
\Beq
\Phi_{\pi}^{CZ}(x,\mu_1) = 6 x(1-x)\left [1 + (5 (1-2 x)^2 -1) 
\left ( \frac{\alpha_s(\mu_1)}{\alpha_s(\mu_{0})} \right )^{50/81} \right ]
\frac{f_{\pi}}{2 \sqrt{6}} \; , 
\label{eq:PiczE}
\Eeq
and, in the modified perturbative approach, the scale $\mu_1$ 
is taken to be $1/b$ \cite {Lietc}.

If we are going to retain the intristic $b$ dependence of the wave functions 
$\Phi$ in the expressions for the matrix elements,
then we are faced with even more uncertainties coming from 
the ambiguity in the form of the wave function $b$-part as well as in the 
values of some new parameters.

The constituent quark model of the wave function associates some Gaussian 
exponential to the $b$-dependent part \cite{DJaKr}, so that
\Beq
\Phi_{\pi}^{as}(x,b) = 6 x(1-x) 4\pi \,\exp \left ( -x (1-x) b^2/(4 a_{as}^2)
\right ) \frac{f_{\pi}}{2 \sqrt{6}}
\label{eq:PiasB}
\Eeq
and
\Beqa
\Phi_{\pi}^{CZ}(x,b,\mu_1) &=& 6 x(1-x)\left [1 + (5 (1-2 x)^2 -1) 
\left ( \frac{\alpha_s(\mu_1)}{\alpha_s(\mu_{0})} \right )^{50/81} \right ]
\nonumber \\
& & 
\cdot \, 4 \pi \, \exp \left ( -x (1-x) b^2/(4 a_{CZ}^2)
\right ) \frac{f_{\pi}}{2 \sqrt{6}} \; , 
\label{eq:PiczB}
\Eeqa
where the pion's transverse parameters $a_{as}$ and $a_{CZ}$ are fixed from 
the $\pi \rightarrow \gamma \gamma $ process to be 
$a_{as} = 0.846 GeV^{-1}$ and $a_{CZ} = 0.655GeV^{-1}$, 
respectively \cite{BoKrS}.

For a $B$-meson wave function there exist a few models \cite{BrJietc}.
We consider two forms that have been proved in the
calculations of various nonleptonic $B$ decays. The first one is \cite{Schl}
\Beq
\Phi_B^{(1)}(x, \vec{k}) = N^{(1)} \left [ C + \frac{m_b^2}{1-x} + 
\frac{\vec{k}^2}{x (1-x)} \right ]^{-2} \; , 
\label{eq:B1k}
\Eeq
whose Fourier transform gives
\Beq
\Phi_B^{(1)}(x,b) = \frac{N^{(1)}}{4 \pi} \frac{b x^2 (1-x)^2}{\sqrt{M_B^2 x + 
C x(1-x)}} \, K_1 \left (\sqrt{M_B^2 x + C x (1-x)} b \right ) \; ,  
\label{eq:B1B}
\Eeq
with the approximation $m_b \simeq M_B = 5.28 GeV$. $K_1$ is the modified 
Bessel function of order one. Neglecting the $b$ dependence leads to 
\Beq
\Phi_B^{(1)}(x) = \frac{N^{(1)}}{16 \pi^2} \frac{x (1-x)^2}{M_B^2 + 
C (1-x)} \; .
\label{eq:B1}
\Eeq
For the constants $N^{(1)}$ and $C$ we have used the fitted parameters 
$N^{(1)}= 604.34 GeV^3$ and $C = -27.5 GeV^2$, which have been proved in other 
calculations \cite{YeLi}.

Another model is the oscillatorlike wave function of Bauer, Stech, and Wirbel
\cite{BSWetc}:
\Beq
\Phi_B^{(2)}(x,b) = \frac{N^{(2)}}{2 \pi} \sqrt{x (1-x)}\; \exp\left 
(- \frac{M_B^2}{2 \omega^2} x^2 \right ) \exp \left ( -\frac{\omega^2}{2} b^2 
\right )\; , 
\label{eq:B2B}
\Eeq
with the constants $N^{(2)}= 156.34 GeV$ and $\omega = 0.4 GeV$. Both wave 
functions, (\ref{eq:B1B}) and (\ref{eq:B2B}), are normalized with $f_B=
200 MeV$.

\narrowtext

The vector-meson wave functions are modeled in the QCD sum-rule 
calculations \cite{CZetc}, \cite{BeCh}. Since the form of the $\phi$-meson 
wave function is still questionable, we have decided to use the asymptotic 
form
\Beq
\Phi_{\phi}(x) = 6 x(1-x) \frac{f_{\phi}}{2 \sqrt{6}}\quad , \quad 
f_{\phi}=0.233 GeV\, , 
\label{eq:Phi}
\Eeq
without including any $b$ dependence. We believe that owing to the lack 
of better experimental data to which transverse parameters can be fixed, an 
unrealistic $b$-dependent part may produce more questionable results than 
by neglecting it.

In order to suppress the soft contributions in the hard scattering 
amplitudes (\ref{eq:aTall}, \ref{eq:aTloop}), we have 
included the Sudakov factors \cite{LiSt}. 
They ensure that the hard scattering amplitude receives contributions only 
from the exchange of hard gluons, suppressing the contributions of soft 
gluons from the large $b$ region. The Sudakov suppression is comprised by 
the hadron wave function redefinition
\Beqa
\Phi_B &\rightarrow& \Phi_B (x_1,b_1)\, \exp( - \displaystyle S_B(t) )\; , 
\nonumber \\
\Phi_{\phi} &\rightarrow& \Phi_{\phi} (x_2,b_2)\, 
\exp( - \displaystyle S_{\phi}(t) )\; , 
\nonumber \\
\Phi_{\pi} &\rightarrow& \Phi_{\pi} (x_3,b_3)\, 
\exp( - \displaystyle S_{\pi}(t) ) \; .
\label{eq:WFsud}
\Eeqa

\widetext

The Sudakov exponentials exhibit the result of all-order resummation of 
double logs 
appearing from the overlap of collinear and soft divergences \cite{BoSt}. 
In our case,
\Beqa
S_B(t) &=& s(x_1 p_1^{-},b_1,t) - 
1/\beta_0 \ln \left (\frac{\ln(t/\Lambda_{QCD})}
{\ln(1/(b_1 \Lambda_{QCD}))} \right ) \; , \nonumber \\
S_{\phi}(t) &=& s(x_2 p_2^{+},b_2,t) + s(((1-x_2) p_2^{+}),b_2,t) - 
1/\beta_0 \ln \left (\frac{\ln(t/\Lambda_{QCD})}
{\ln(1/(b_2 \Lambda_{QCD}))} \right )  \; , \nonumber \\
S_{\pi}(t) &=& s(x_3 p_3^{-},b_3,t) + s(((1-x_3) p_3^{-}),b_3,t) - 
1/\beta_0 \ln \left (\frac{\ln(
t/\Lambda_{QCD})} {\ln(1/(b_3 \Lambda_{QCD}))} \right )\; , 
\label{eq:Sexp}
\Eeqa
where $\beta_0 = (33 - 2 n_f)/12$ and $n_f = 4$. 
For $\Lambda_{QCD}$ we have used the value $\Lambda_{QCD} = 0.2 GeV$ 
throughout the paper. 
The last term in the above expressions accounts for the renormalization from 
the IR scale $1/b$ to the some renormalization scale $t$, which we are going 
to take to be one of the scales from (\ref{eq:texp}), depending on the 
diagram considered.

\narrowtext

The full expressions for the Sudakov functions $s(x_i,b_i,t)$, together with 
the usual approximations used in a numerical treatment, can be found in 
\cite{WuYeLi}.

Note that we have also associated the Sudakov function $s(x_1 p_1^{-}, b_1,t)$ 
with the light antiquark of the B meson.
The heavy $b$ quark, having a finite mass, does not produce collinear 
divergences and its Sudakov function is zero.

Use of the above mentioned diversity of the wave functions would certainly 
diminished the capability of perturbative calculations for giving reliable 
predictions for the $B^{-} \rightarrow \pi^{-} \phi$ branching ratio and CP
asymmetry, having in mind that
the effects of the large reduction of the results
owing to the intrinsic $b$-dependence in the wave functions
as well as the large difference in the predictions depending on the 
$B$ and $\pi$ meson wave function
employed, has already been observed
in other perturbative calculations \cite{LiYu}. We have checked that
this is also the case in the calculation of the $B \rightarrow \pi \phi$ decay. 

Owing to the specific character of the $B \rightarrow \pi \phi$ decay governed 
by the $b \rightarrow d s \overline{s}$ transition where the strange 
quark-antiquark pair has to form the final $\phi$-meson state, 
we can assume that 
the $B \rightarrow \pi \phi$ process is determined predominantly by 
the $B \rightarrow \pi$ transition at the energy $p^2 = M_{\phi}^2$. 

Therefore, we can try to make a selection among the wave functions by comparing 
the results for the $B \rightarrow \pi$ transition form factor obtained from 
the QCD sum rule \cite{QCD} and lattice calculations \cite{latt} summarized in 
\Beq
F_{+}^{B \rightarrow \pi}(0) = 0.25-0.35\; , 
\label{eq:ffQCD}
\Eeq
with those estimated in our modified perturbative approach. 

\widetext

The expression for 
the form factor in the perturbative approach has the form
\Beqa
F_{+}^{B \rightarrow \pi}(\eta) &=& \frac{C_F}{2} M_B^2  \int dx_1 dx_3 \,
\int b_1 db_1\, b_3 db_3 \Phi_{B}(x_1,b_1) \Phi_{\pi}^{\ast}(x_3,b_3)\,
\nonumber \\
&\times& \left \{
\frac{\alpha_s(t_A)}{\pi} H(D_G,D_b,b_1,b_3)\,
[1+x_3 \eta ]\, e^{- \displaystyle (S_B(t_A) +
S_{\pi}(t_A) )} \right . \nonumber \\
& & \left .  +\frac{\alpha_s(t_B)}{\pi} H(D_G,D_{q_1},b_3,b_1)\,
[ -x_1 (1-\eta)] 
\, e^{- \displaystyle (S_B(t_B) + S_{\pi}(t_B) )}    \right \} \; ,
\label{eq:ff}
\Eeqa
which can be easily recognized in the expression for the factorizable part of 
the $B \rightarrow \pi \phi$ decay, Eq. (\ref{eq:aTall}). 
The parameter $\eta$ is the fraction of the energy of the $\pi$ meson and at 
the momentum transfer $p^2 = 0$ or $p^2 = M_{\phi}^2$ we have $\eta =1$ or 
$\eta = 1-M_{\phi}^2/M_{B}^2 = 1 - r^2$, respectively.

\narrowtext

Estimating the $B \rightarrow \pi$ transition form factor at the momentum 
transfer $p^2=0$ using different forms of the $B$ and $\pi$ meson wave 
functions 
taken from above, we achieve predictions which are far from the values 
obtained in the QCD sum rule and lattice calculations (\ref{eq:ffQCD}), 
except if we 
assume the oscillatorlike model for the $B$ meson wave function 
$\Phi_B^{(2)}(x)$, (\ref{eq:B2B}), and the CZ type of the pion wave function 
(\ref{eq:PiczE}), both being intrinsic 
b-independent.  
Our predicted value for the $B \rightarrow \pi$ form factor obtained with 
these wave functions is 
\Beq
F_{pert}^{B \rightarrow \pi}(0) =  0.282 \; .
\label{eq:ffper}
\Eeq

Both, $\Phi_B^{(2)}(x)$ and $\Phi_{\pi}^{CZ}(x, \mu_{1})$ are more end-point 
concentrated wave functions than their alternative forms, 
$\Phi_B^{(1)}(x)$ (\ref{eq:B1}) and $\Phi_{\pi}^{as}(x)$ (\ref{eq:Pias}), 
respectively. This 
indicates a need for the enhancement of the soft contributions in order to 
match the predictions (\ref{eq:ffQCD}) for the $B \rightarrow \pi$ form factor 
esimated by nonperturbative methods.
 
Comparable calculations of the
$B \rightarrow \pi$ form factor in the modified perturbative 
approach have also been performed in \cite{DJaKr}, \cite{LiYu} 
and, similarly, the results obtained have exibited strong dependence on 
the mesonic wave functions used, 
confirming that the wave functions represent the weakest point in the 
calculation of $B$-meson decays in the perturbative approach.

\section{Numerical results and discussions}

Now we are going to discuss the branching ratio $BR$ and the 
CP asymmetry in the $B^{-} \rightarrow \pi^{-} \phi$ 
decay numerically.

The decay rate is given as
\Beq
\Gamma (B^{-} \rightarrow \pi^{-} \phi) = \frac{1}{16 \pi} 
\frac{\lambda^{1/2} (M_B, M_{\phi},0)}{M_B^3} |{\cal M}|^2 \; , 
\label{eq:rate}
\Eeq
where $\lambda^{1/2}(M_B,M_{\phi},0) = M_B^2 (1-r^2)$ and the total amplitude 
${\cal M}$ is given by (\ref{eq:Mamp}). CP asymmetry 
in terms of the ${\cal A}_u$ and ${\cal A}_c$ amplitudes, Eq.(\ref{eq:Mamp}), reads
\Beq
a_{CP} = \frac{-2 V_c Im(V_u) Im({\cal A}_u {\cal A}_c^{\ast})}{( |V_u|^2 
|{\cal A}_u|^2 + |V_c|^2 
|{\cal A}_c|^2 + 2 V_c Re(V_u) Re({\cal A}_u {\cal A}_c^{\ast}) )}\; . 
\label{eq:aCPexp}
\Eeq

The products of the CKM matrix elements may be written in the Wolfenstein 
parametrization as
\Beqa
V_u &=& V_{ud}^{\ast} V_{ub} = A \lambda^3 (1-\lambda^2/2) (\rho- i \eta) 
 \equiv A \lambda^3 (\overline{\rho} - i\,\overline{\eta})\; , \nonumber \\
V_c &=& V_{cd}^{\ast} V_{cb} = - A \lambda^3 \; .
\Eeqa

We use the following values of the parameters $\overline{\rho}$ and 
$\overline{\eta}$:
\Beq
\overline{\rho} = 0.16\; , \;\;\; \overline{\eta} = 0.33 \; , 
\label{eq:rhoeta}
\Eeq
which correspond to their central values obtained by the unitarity fit
\cite{Parodi}. Since recent measurements disfavor the negative values for 
the $\rho$ parameter \cite{exp}, the CP asymmetries will be presented in 
figures by taking $\overline{\rho}$ in the range
\Beq
0 \le \overline{\rho} \le 0.25
\Eeq
and using the central value for the $\overline{\eta}$ parameter from 
(\ref{eq:rhoeta}), 
the minimum and the maximum allowed values \cite{Parodi}
\Beqa
\overline{\eta} = 0.27 \; , \nonumber \\
\overline{\eta} = 0.38 \; ,   
\label{eq:eta}
\Eeqa
respectively.  The other 
Wolfenstein CKM parameters used are $A = 0.823$ and $\lambda = 0.2196$.

Following Ref. \cite{Alietc}, we are going to take the constituent
quark masses in the loop expressions, Eqs.(\ref{eq:dG}) and (\ref{eq:aTloop}), 
with particular values $m_u = 0.2 GeV$ and $m_c = 1.5 GeV$. 

For the Wilson scale-independent coefficients at the renormalization 
scale $\mu = m_b = 4.8 GeV$ ($\alpha_s(M_Z) = 0.118$, $\alpha_{\rm em}(M_Z) = 
1/128$) we take \cite{KrPa}
\Beqa
{\overline c}_{1} &=& -0.324 \; , \;\;\;\;\;\;
{\overline c}_{2} = 1.15 \; , \nonumber \\
{\overline c}_{3} &=& 0.017 \; , \;\;\;\;\;\; 
{\overline c}_{4} = -0.038  \; , \nonumber \\
{\overline c}_{5} &=& 0.011 \; , \;\;\;\;\;\;
{\overline c}_{6} = -0.047  \; , \nonumber \\
{\overline c}_{7} &=& -1.05\cdot 10^{-5} \; , \;\;\;\;\;\;
{\overline c}_{8} = -3.84\cdot 10^{-4}  \; , \nonumber \\
{\overline c}_{9} &=& -0.0101 \; , \;\;\;\;\;\;
{\overline c}_{10} = 1.96\cdot 10^{-3} \; . 
\label{eq:WC1}
\Eeqa

One can note from expression (\ref{eq:aCPexp}) that
some absorptive part in the amplitude is essential for nonvanishing 
CP asymmetry. 

As in the BSW factorization approach, the necessary absorptive part 
comes from 
the cut in the penguinlike diagrams in Fig.\ref{f:main}(c), 
residing in the term $\Delta G(m_q^2,M_{\phi}^2, 
\mu^2)$. From 
expression (\ref{eq:dG}) it is easy to see that the absorptive part is 
developed for the virtual photon momentum $q$ such that $q^2 \ge 4 m_q^2$, 
$q = u,c$. 
Owing to the specific momentum distributions in the $B^{-} \rightarrow 
\pi^{-} \phi$ process, the imaginary part emerges only from 
the diagram with a $u$ quark in the loop. In numerical calculations we use 
the approximation of (\ref{eq:dG}), 
\Beq
\Delta G_{app}(m_q^2,M_{\phi}^2,\mu^2) = \frac{2}{3} \left [\frac{5}{3} + 
\frac{4}{z} + 
(1 + \frac{2}{z}) R(z) - \ln \frac{m_q^2}{\mu^2} \right ] \; , 
\Eeq
where by defining $a = \sqrt{|1 - 4/z|}$, we have
\Beqa
R(z) = \left \{ \begin{array}{l}
-a \pi + 2 a \arctan(a)\quad , \quad  z = ({M_{\phi}}/{m_c})^2 \\
\\
i a \pi + a \ln \displaystyle \frac{1-a}{1+a} \quad , \quad  
z = ({M_{\phi}}/{m_u})^2
\end{array} \right . \;.
\Eeqa

\widetext

In addition, in the perturbative approach there are absorptive parts 
connected with the cuts in the propagators of virtual partons in each of 
the diagrams in Fig. \ref{f:main}.
The expression for $H$, (\ref{eq:FTh}), can develop the imaginary part for 
some of the values
of the fractions $x_i$ in the integration for which the denominators of the 
gluon or quark 
propagators under the square root become negative
 (see Eqs.(\ref{eq:prop1}, \ref{eq:prop2})). 
In this case we take
\Beq
K_0(i y b) = \frac{i \pi}{2} H_0^{(1)} (y b)
\Eeq
and
\Beq
f(i y,b_1,b_2) = \frac{i \pi}{2} \left [ \Theta (b_1-b_2)
H_0^{(1)}(y b_1) J_0(y b_2)
+ \Theta (b_2-b_1) H_0^{(1)}(y b_2) J_0(y b_1) \right ] \; .
\Eeq

%
\begin{table}
\caption{ Branching ratios for the
$B^{-} \rightarrow \pi^{-} \phi$ decay calculated for different
penguin contributions, by taking only the factorizable parts or the complete
expression into account. The first column contains the predictions
obtained in the BSW factorization approach by using $\langle q^2 \rangle =
M_{\phi}^2$ (see text). Columns I and II give predictions
calculated in the modified perturbative approach by employing the meson wave
functions $\Phi_B^{(2)}(x)$, Eq.(\protect\ref{eq:B2}) and
$\Phi_{\pi}^{CZ}(x, 1/b)$,  Eq.(\protect\ref{eq:PiczE}), and by using
the Wilson coefficients ${\overline c}_k(m_b)$ (\protect\ref{eq:WC1})
in column I and ${\overline c}^{(0)}_k(t)$ (\protect\ref{eq:WC2}) in
column II. The CKM  parameters used are $\overline{\rho} = 0.16$ and
$\overline{\eta} = 0.33$.}
\begin{tabular}{lccc}
Penguin contributions &
\multicolumn{3}{c}{BR}  \\
  \cline{2-4} &
BSW & I & II \\ \hline
\\
QCD-factorizable & $0.20\cdot 10^{-10}$ & $0.14\cdot 10^{-10}$
& $1.06\cdot 10^{-10}$ \\
QCD-all & -& $2.51\cdot 10^{-10}$  & $0.73\cdot 10^{-10}$ \\
\\
QCD+QED-factorizable & $0.34\cdot 10^{-8}$ & $0.38\cdot 10^{-8}$
& $0.89\cdot 10^{-8}$ \\
QCD+QED-all &- & $0.44\cdot 10^{-8}$  & $0.85\cdot 10^{-8}$ \\
\\
\end{tabular}
\label{t:rBR}
\end{table}

Having selected the B meson wave function 
$\Phi_B^{(2)}(x)$, (\ref{eq:B2B}), in the preceding section:
\Beq
\Phi_B^{(2)}(x) = \frac{1}{4 \pi} \Phi_B^{(2)}(x,0) = 
\frac{N^{(2)}}{8 \pi^2} \sqrt{x (1-x)}\; \exp\left
(- \frac{M_B^2}{2 \omega^2} x^2 \right ) \; ,
\label{eq:B2}
\Eeq
and the $\Phi_{\pi}^{CZ}(x,1/b)$ (\ref{eq:PiczE}) for the $\pi$ meson 
wave function, we can 
now continue along the lines developed in the preceding sections and 
give reliable 
predictions for the $B^{-} \rightarrow \pi^{-} \phi$ branching ratio and CP
asymmetry in the modified perturbative approach, using the NLO weak Hamiltonian. 
\narrowtext

The results are presented in Tables I and II, together with the predictions 
estimated in the BSW factorization approach, both being calculated with 
the preferred values of the CKM 
parameters, $\overline{\rho} = 0.16$ and $\overline{\eta} = 0.33$. 

%
%
\begin{table}
\caption{ CP asymmetries for the
$B^{-} \rightarrow \pi^{-} \phi$ decay calculated for the QCD and QED
penguin contributions together,
by taking only the factorizable parts or the complete
expression into account. The first column contains the predictions
obtained in the BSW factorization approach by using $\langle q^2 \rangle =
M_{\phi}^2$ (see text). Columns I and II give predictions
calculated in the modified perturbative approach by employing the meson wave
functions $\Phi_B^{(2)}(x)$, Eq.(\protect\ref{eq:B2}) and
$\Phi_{\pi}^{CZ}(x, 1/b)$,  Eq.(\protect\ref{eq:PiczE}), and by using
the Wilson coefficients ${\overline c}_k(m_b)$ (\protect\ref{eq:WC1})
in column I and ${\overline c}^{(0)}_k(t)$ (\protect\ref{eq:WC2}) in
column II. The CKM  parameters used are $\overline{\rho} = 0.16$ and
$\overline{\eta} = 0.33$.}
\begin{tabular}{lccc}
Penguin contributions &
\multicolumn{3}{c}{$a_{CP}/10^{-2}$} \\
  \cline{2-4} &
BSW & I & II \\ \hline
\\
QCD+QED-factorizable  & -1.9 & 14.6 & 15.4  \\
QCD+QED-all  &  - & 16.1 & 16.3  \\
\\
\end{tabular}
\label{t:rCP}
\end{table}

Calculations of the $B^{-} \rightarrow \pi^{-} \phi$ branching ratio and 
asymmetry in the BSW factorization approach have been performed by many authors
\cite{Fl}-\cite{ALI}. In order to be able to clearly assign the role of 
nonfactorizable 
contributions in the decay, we have recalculated the BSW factorization 
predictions using our values of the Wilson coefficients $\overline{c}_i$ 
(\ref{eq:WC1}) and 
the CKM parameters $\overline{\rho}$ and $\overline{\eta}$. 

The decay amplitude in the BSW approach can be directly compared with the 
complete expression for the $B \rightarrow \pi \phi$ amplitude given by Eq.(
\ref{eq:Mamp}) by neglecting nonfactorizable parts and numerically suppressed  
contributions emerging from the diagrams in Fig. 3(d). The matrix element 
in the strict factorization approach is propotional to the 
$F^{B \rightarrow \pi}(M_{\phi}^2)$ form factor which we calculate 
in the single-pole approximation as
\Beq
F_{BSW}^{B \rightarrow \pi}(M_{\phi}^2) = \frac{F_{BSW}^{B \rightarrow \pi}(0)}
{1- M_{\phi}^2/M_{B\,\pi}^2(1^{-})} \; , 
\label{eq:ffBSW}
\Eeq
where $M_{B\,\pi}^2(1^{-}) = 5.32 GeV$ and $F_{BSW}^{B \rightarrow \pi}(0) = 
0.33$ \cite{BSW}. 
It is worth mentioning that the prediction for the $B \rightarrow \pi$ 
form factor estimated in the perturbative approach, Eq.(\ref{eq:ffper}), is 
somewhat smaller than the $F_{BSW}^{B \rightarrow \pi}(0)$ value. 

All results estimated in the factorization approach are obtained by taking 
the virtual photon momentum squared equal to $q^2 = M_{\phi}^2$. We have 
already stated that, in general, information about the $q^2$-value is 
lost by factorizing hadronic matrix elements, except in the strict 
factorization, when nonfactorizable and/or 
strong final-state interactions are neglected. Therefore, 
considering possible nonfactorizable or final-state corrections, 
after the factorization procedure, the 
$q^2$ is usually considered as a free parameter,  
whose average value is constrained by some simple, general kinematical 
reasons to be \cite{Tramp}
\Beq
m_b^2/4 \le  \langle q^2 \rangle \le m^2_b/2 \; ,
\label{eq:q2}
\Eeq 
and usually assumed to be valid for all nonleptonic heavy-to-light transitions.

The dependence of the branching ratio and CP asymmetry on the 
$\langle q^2 \rangle$ as a function of the $\overline{\rho}$ 
CKM-parameter is shown in 
Figs. \ref{f:BSW}. The branching ratio appears to be practically independent 
of the value of the $\langle q^2 \rangle$. Such behavior is due to the 
cancellation which occurs between the Wilson coefficients multiplying 
the one-loop QED penguinlike matrix element, (\ref{eq:Mamp}), because of the 
relation
\Beq
3 \overline{c}_1 (m_b) \simeq - \overline{c}_2 (m_b) \; .
\Eeq

\begin{figure}
 \centerline{  \epsfig{file=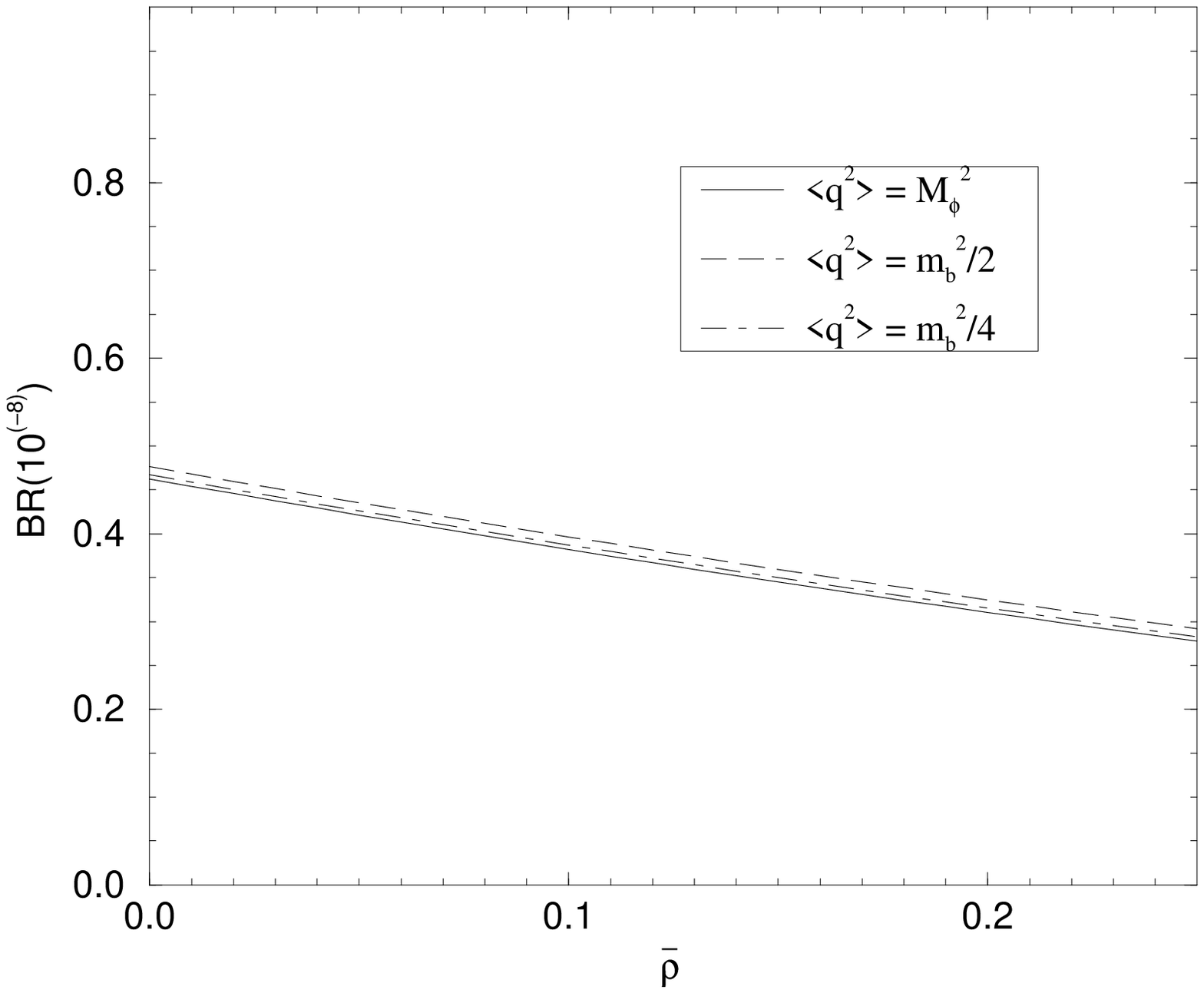,height=7.5cm,width=8cm, silent=}
   \epsfig{file=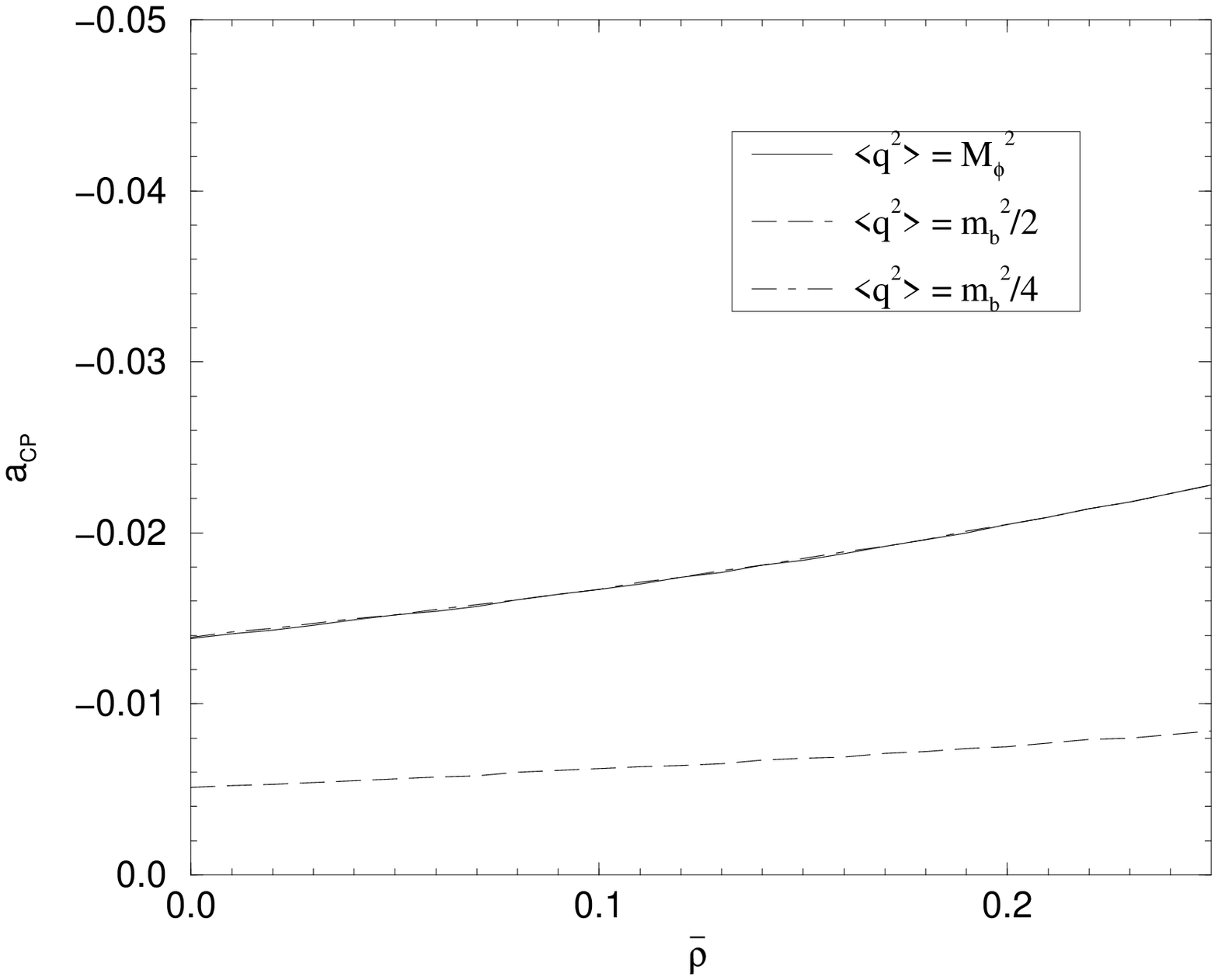,height=7.5cm,width=9cm,silent=} }
\caption{ Branching ratio and CP asymmetry in the
$B^{-} \rightarrow \pi^{-} \phi$ decay calculated in the BSW factorization
approach as a function of the CKM parameter $\overline{\rho}$ and for
the central value of $\overline{\eta} = 0.33$.
The solid, long dashed, and dot-dashed
lines denote predictions obtained by taking $\langle q^2 \rangle =
M_{\phi}^2$, $\langle q^2 \rangle = m_b^2/2$, and
$\langle q^2 \rangle = m_b^2/4$, respectively (see text). }
\label{f:BSW}
\end{figure}

On the contrary, CP asymmetry exibits a large reduction of up to $70 \%$ if 
higher $\langle q^2 \rangle $-values are taken.
 
The results in column II in Tables I and II are obtained by calculation
inspired by the papers of
Li and collaborators \cite{YeLi}, \cite{LiColl}, in which the $\mu$
scale-setting ambiguity of
Wilson coefficients is moderated by applying the three-scale
factorization theorem.
Their theorem keeps trace of all three scales characterizing the
nonleptonic weak decay, the W-boson mass $M_W$, the
typical scale $t$ of the process, and the hadronic scale $\sim \Lambda_{QCD}$,
and proves for the leading-order weak Hamiltonian that Wilson coefficients
should be taken at the scale $t$, a typical scale
in a particular decay. The matrix elements of the operators ${\cal O}_k$
and
Wilson coefficients are then both calculated at the same scale. 
The scale is determined by the dynamics of the process, contrary to
the arbitrary renormalization scale $\mu$ taken to be a constant, $m_b$, 
for the ${\overline c}_k$ coefficients, (\ref{eq:WC1}).

\widetext

Under the assumption that the three-scale factorization theorem is also valid
for the NLO weak Hamiltonian (\ref{eq:ewH1}) we have taken the explicit form of
the Wilson coefficients, calculated directly at $M_W$ and
then rescaled to some lower scale $t$ \cite{Fl}:
\Beqa
{\overline c}^{(0)}_1 (t) &=& {\cal O}(\alpha_s(t)) +
{\cal O}(\alpha_{\rm em}) \; , \nonumber \\
{\overline c}^{(0)}_2 (t) &=&  1 + {\cal O}(\alpha_s(t)) +
{\cal O}(\alpha_{\rm em}) \; , \nonumber \\
{\overline c}^{(0)}_3 (t)&=& -\frac{\alpha_s(t)}{24 \pi} \left [ E_0(x_t) +
\frac{2}{3} \log \frac{t^2}{M_W^2} - \frac{10}{9} \right ]
+ \frac{\alpha_{\rm em}}{6 \pi} \frac{1}{\sin^2 \Theta_W} \left [ 2 B_0(x_t) +
C_0(x_t) \right ] \; , \nonumber \\
{\overline c}^{(0)}_4 (t) &=& \frac{\alpha_s(t)}{8 \pi} \left [ E_0(x_t) +
\frac{2}{3} \log \frac{t^2}{M_W^2} - \frac{10}{9} \right ] \; , \nonumber \\
{\overline c}^{(0)}_5 (t) &=& -\frac{\alpha_s(t)}{24 \pi} \left [ E_0(x_t) +
\frac{2}{3} \log \frac{t^2}{M_W^2} - \frac{10}{9} \right ] \; , \nonumber \\
{\overline c}^{(0)}_6 (t) &=& \frac{\alpha_s(t)}{8 \pi} \left [ E_0(x_t) +
\frac{2}{3} \log \frac{t^2}{M_W^2} - \frac{10}{9} \right ] \; , \nonumber \\
{\overline c}^{(0)}_7 (t) &=& \frac{\alpha_{\rm em}}{6 \pi} \left [
4 C_0(x_t) + D_0(x_t) +
\frac{4}{9} \log \frac{t^2}{M_W^2} - \frac{20}{27} \right ] \; , \nonumber \\
{\overline c}^{(0)}_8 (t) &=& 0 \; , \nonumber \\
{\overline c}^{(0)}_9 (t)&=& \frac{\alpha_{\rm em}}{6 \pi} \left [ 4 C_0(x_t)
 + D_0(x_t) + \frac{4}{9} \log \frac{t^2}{M_W^2} - \frac{20}{27} \right ]
+ \frac{\alpha_{\rm em}}{3 \pi} \frac{1}{\sin^2 \Theta_W} \left [ 5 B_0(x_t)
- 2 C_0(x_t) \right ] \; , \nonumber \\
{\overline c}^{(0)}_{10} (t) &=& 0 \; ,
\label{eq:WC2}
\Eeqa
where $x_t = m_t^2/M_W^2$.
The functions $B_0$, $C_0$, $D_0$, and $E_0$ are the Inami-Lim functions
\cite{InLim}.

\narrowtext

These coefficients are only an approximation of the Wilson coefficients
${\overline c}_{k}(\mu = m_b)$, (\ref{eq:WC1}) obtained by performing the
renormalization-group analysis \cite{BuJaLW} and
used throughout the paper, but we hope that
possible uncertainties
involved in the calculation by using the coefficients in (\ref{eq:WC2})
are covered within the accuracy of our model. In the numerical estimates 
we have also taken into account 
${\cal O}(\alpha_s)$ corrections in ${\overline c}^{(0)}_1$ and
${\overline c}^{(0)}_2$ in order to have the proper
${\cal O}(\alpha_s \alpha_{\rm em})$ calculation.

By taking the Wilson coefficients ${\overline c}_k^{(0)}$ at one of
the scales
(\ref{eq:texp}), depending on the diagram involved as a contribution of the
operator ${\cal O}_k$, we obtain the results given in column II in Tables I 
and II. 
The results are estimated again with the selected 
wave functions $\Phi_B^{(2)}(x)$ and $\Phi_{\pi}^{CZ}(x,1/b)$.

Let us now discuss the results from Tables and emphasize 
their general characteristics.

One can note that the $B^{-} \rightarrow \pi^{-} \phi$ 
process is clearly dominated by the EW penguin contributions, in both the 
factorization and the perturbative approaches and the predicted branching ratio 
for the $B^{-} \rightarrow \pi^{-} \phi$ decay is of order 
${\cal O}(10^{-8})$. 

It is obvious that EW nonfactorizable contributions are small, being directly
proportional to the small Wilson coefficients
${\overline c}_8$ and ${\overline c}_{10}$. For nonfactorizable
contributions of
QCD penguin operators there is no such
apparent reason, because the Wilson coefficients multiplying the operators 
${\overline c}_4$
and ${\overline c}_6$, are in absolute magnitude even larger
than the coefficient ${\overline c}_9$, which dominates the $B^{-} \rightarrow
\pi^{-} \phi$ decay (\ref{eq:WC1}). The
influence of the QCD nonfactorizable contributions is noticeable, specially 
in the perturbative results based on the Wilson coefficients taken from 
(\ref{eq:WC1}) and represented in column I. For this case, by comparing 
the third and fourth rows in Table I, we can see that nonfactorizabile 
corrections can account for some $14\%$ of the final result. 
However, after taking the Wilson coefficients in the convolution with the 
hadronic matrix elements at the same scale, 
 as it is done by obtaining the results in column II, nonfactorizable 
contributions become negative, and small, and can be considered negligible, 
lowering the final result by some $4 \%$. Negligible nonfactorizable 
corrections in this model indicate that, by a suitably chosen scale which 
truly makes the product of the Wilson coefficients and the hadronic matrix 
elements scale independent, it is possible to account for the almost strict 
factorization in the $B^{-} \rightarrow \pi^{-} \phi$ decay, which 
would be naively expected by the "color transparency argument" \cite{NS}. 

Further general behavior of the results from column II can be summarized in 
the statement that the branching ratios calculated using the Wilson 
coefficients ${\overline c}_k^{(0)} (t)$ 
are enlarged by some factor two in comparison with the 
estimations obtained by using  ${\overline c}_k(\mu = m_b)$.
In addition, CP asymmetries are predicted to be about $16\%$, similarly 
as in the model in column I, where ${\overline c}_k(m_b)$ are used.

The predicted asymmetries are much larger than those obtained from the 
BSW factorization model, and they are of an opposite sign. 
The reason for such an enlargement of CP asymmetry are absorptive 
contributions due to the on-shell effects in the propagators of 
virtual partons appearing in the perturbative calculation. They are also 
present in the factorizable amplitude $\langle T_{fact} \rangle $ 
which multiplies the penguin loop. Neglecting of the imaginary parts coming 
from the on-shell effects in the propagators would give predictions for 
CP asymmetry comparable with that obtained in the factorization approach.

Figures \ref{f:per} show the impact of 
different choices of CKM parameters on our predictions for the 
branching ratios and CP asymmetry in the 
$B^{-} \rightarrow \pi^{-} \phi$ decay. The predicted asymmetries 
calculated by using the Wilson coefficients ${\overline c}_k(m_b)$ 
and ${\overline c}_k^{(0)} (t)$ are almost the same. Therefore, we show 
explicitly only the CP asymmetry obtained by using ${\overline c}_k^{(0)} (t)$ 
coefficients. 
One can note that the predicited CP asymmetry can be enlarged up to 
$22\%$.    

\begin{figure}
    \centerline{ \epsfig{file=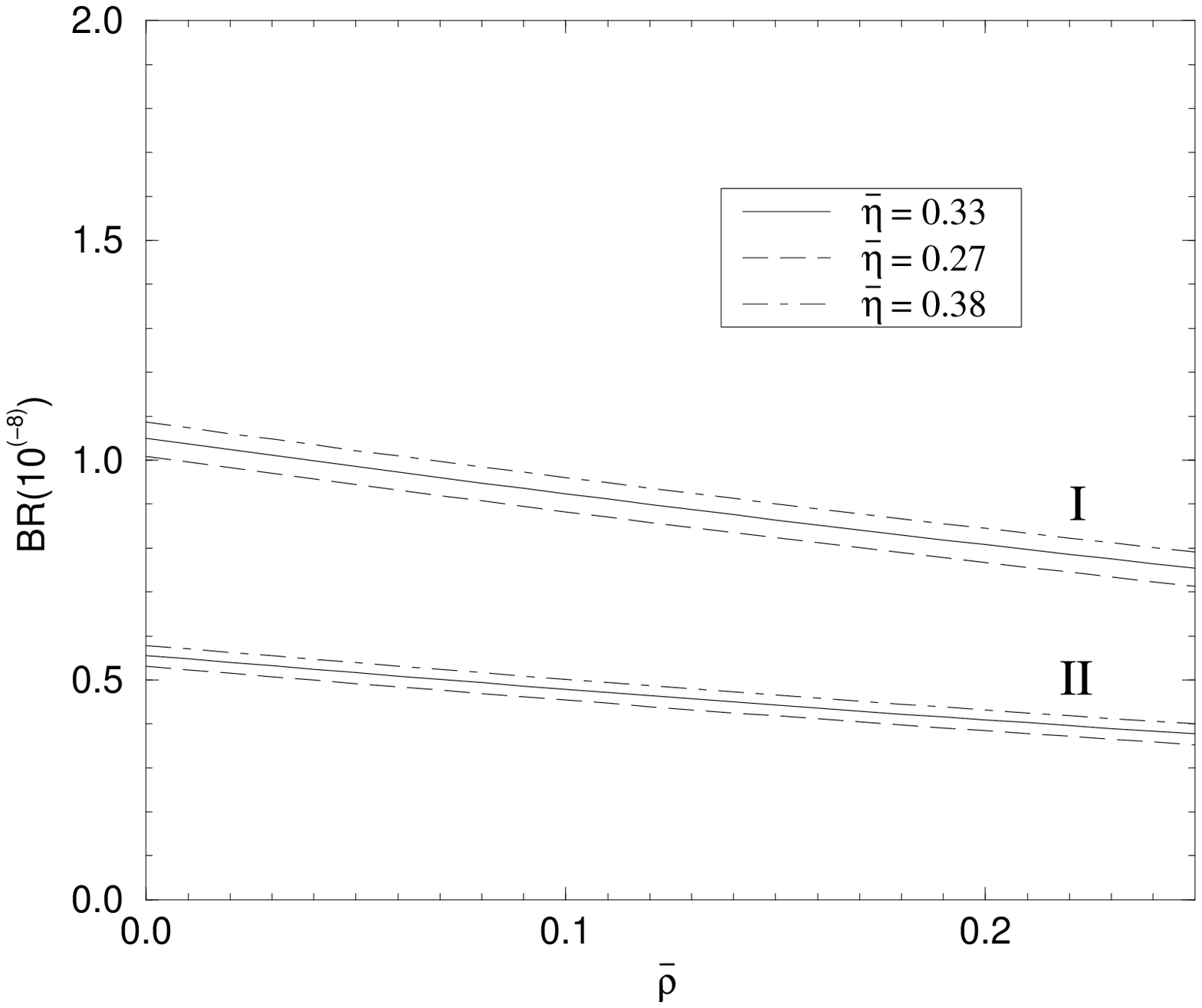,height=7.5cm,width=8cm, silent=}
     \epsfig{file=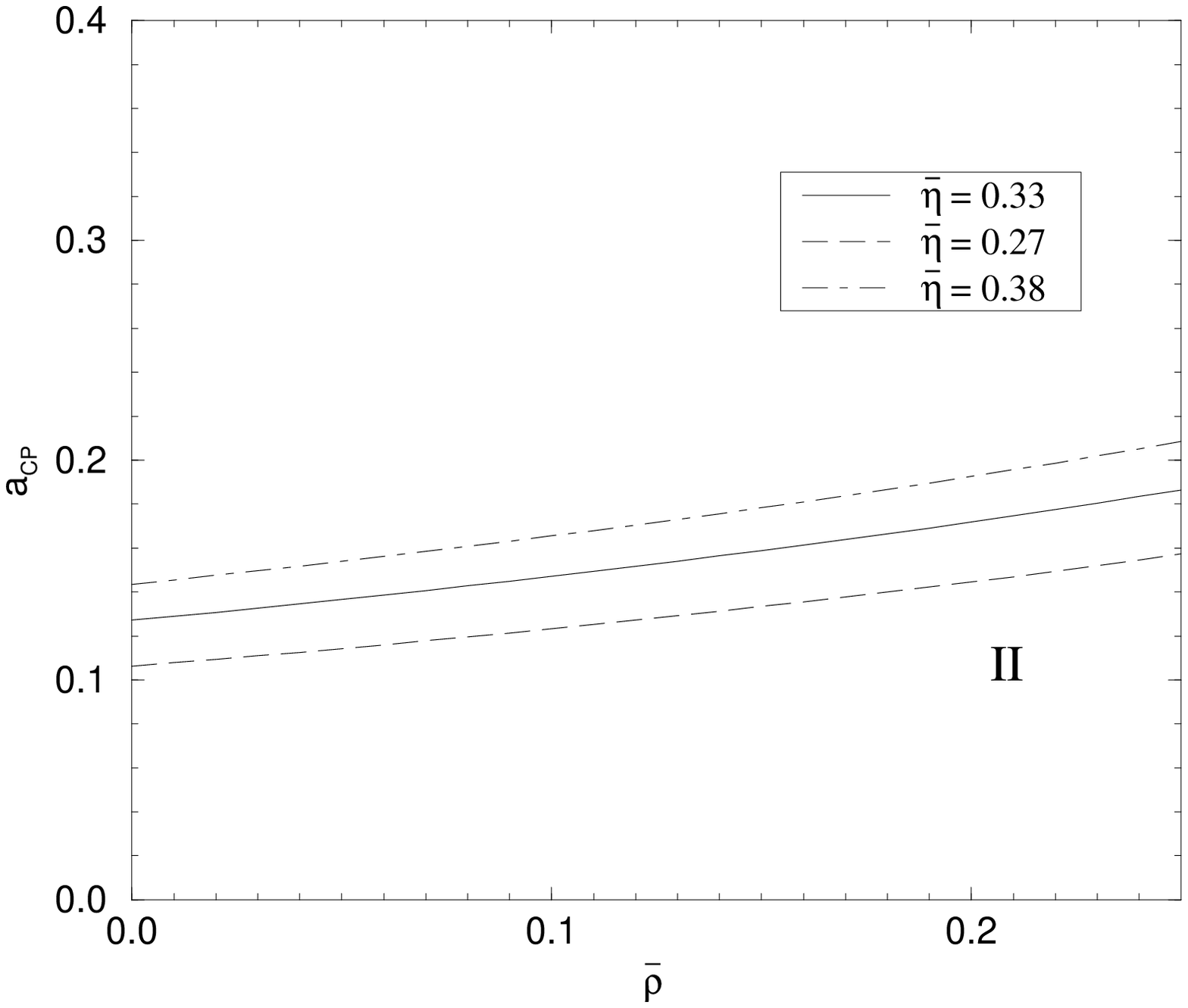,height=7.5cm,width=8cm,silent=} }
\caption{ Branching ratio and CP asymmetry in the
$B^{-} \rightarrow \pi^{-} \phi$ decay calculated in the modified
perturbative
approach as a function of the CKM parameter $\overline{\rho}$.
The solid, long dashed, and dot-dashed
lines correspond to the values of the CKM parameter
$\overline{\eta} = 0.33$, $\overline{\eta} = 0.27$, and
$\overline{\eta} = 0.38$, respectively.
Predictions obtained by using the Wilson coefficients
${\overline c}_k(m_b)$ (\protect\ref{eq:WC1}) and
${\overline c}^{(0)}_k(t)$ (\protect\ref{eq:WC2}) are denoted by labels
I and II, respectively. }
\label{f:per}
\end{figure}

\section{Conclusions}

In this paper we have calculated the branching ratio and CP asymmetry 
of the penguin-induced $B^{-} \rightarrow \pi^{-} \phi$ decay in the 
modified perturbative approach by applying the NLO effective weak Hamiltonian. 
Working in the framework of the modified perturbative approach we have included 
the transverse momentum dependence and the Sudakov form factors. 
The modified perturbative approach also enables us to calculate 
nonfactorizable contributions. 

We have used the $B \rightarrow \pi$ transition form factor to 
select mesonic wave functions by comparing our result with the predictions 
estimated in the QCD sum rule and lattice calculations. 
The comparable prediction has been  
obtained only for the intrinsic b-independent, more end-point concentrated 
wave functions for both $B$ and $\pi$ mesons, $\Phi_{B}^{(2)}(x)$ (\ref{eq:B2}) 
and $\Phi_{\pi}^{CZ}(x,1/b) $ (\ref{eq:PiczE}), respectively. 

Using the NLO weak Hamiltonian and the selected wave functions 
we have first worked with the 
renormalization 
scheme-independent coefficients and have been able to calculate the EW penguin 
contributions properly, proving their dominance in the 
$B^{-} \rightarrow \pi^{-} \phi$ decay. 

In addition, we have examined the assumption of taking the Wilson 
coefficients
to be convolution functions in the starting factorization formula
(\ref{eq:ME})
instead of taking them as constants at some arbitrary scale $\mu$.
The Wilson coefficients then enter into the factorization formula in the
convolution
with the matrix elements at the same scale $t$, typical of the process
and that resolves the problem of different renormalization scales for the
short-distance part (Wilson coefficients) and the long-distance part (matrix
elements of the four-quark operators) in the amplitude of the weak Hamiltonian.
Estimations based on this assumption have produced the branching ratios about
factor two larger than those calculated 
with the conventional Wilson coefficients. 

Besides, if the Wilson
coefficients are considered to be functions of the scale,
the same one
which appears in the hadronic matrix elements, then the nonfactorizable
QCD penguin contributions appear to be negligible, as is the case with 
the obviously very small nonfactorizable
contributions of the EW penguin operators. 

Therefore, our results for the branching ratio appear to be in agreement
with previous calculations performed
in the BSW factorization approach, predicting the branching ratio to be of
order ${\cal O}(10^{-8})$, dominated by the EW penguins. On the other side,
the predicted CP asymmetry differs a lot from that estimated in
the BSW factorization approach, being as large as $16\%$ and having 
an opposite sign for the 
preferred values of the CKM parameters $\overline{\rho} = 0.16$ and 
$\overline{\eta} = 0.33$. The large 
CP asymmetry estimated in the perturbative approach is the result of 
large on-shell
effects of the virtual propagators involved in the calculation. 

The strong reduction of the results obtained with the intrinsic b-dependence 
of the wave functions indicates that 
mesonic wave functions still needs further investigations. 
Presently, $B$-meson wave functions suffer from uncertaintes involved in 
the models from which they are derived as well as from uncertaintes 
coming from the fit to experimental data, and ask for a more refined 
treatment in their derivation. 

Provided that the $B$-meson wave function could be better determined, 
the formalism of this paper may be successfully applied to similar 
penguin-induced decays, 
of which $B^{-} \rightarrow K^{-} \phi$ and $B^{-} \rightarrow \omega \phi$ 
are 
particularly interesting owing to recent experimental measurements and their 
role in the determination of the values of some CKM matrix elements. 
These topics will be the subject of our future investigations.

\acknowledgments

  The author would like to thank Hsiang-nan Li for helpful discussions.
  This work was supported by the Ministry of Science and Technology
  of the Republic of Croatia under Contract
  No. 00980102.

\widetext

\appendix
\section{}

The results of the calculation of diagrams $E$ and $F$, shown in Fig. 
\ref{f:main}(d), are given here explicitly. 
In the calculation we have neglected the 
transverse momenta in the loops. The general expression for the contribution 
of the diagrams can be written as
\Beq
T_q^{i} = C^{i} + I_q^{i} + H_q^{i} + G_q^{i} \quad, \quad\quad i = E,F \; .
\Eeq
The results are presented as integrals over Feynman parameters 
and the particular contributions are found to be 
\Beqa
C^E &=& (-) \frac{1}{6} [ 2 (1-r^2)(1-2 x_1) \epsilon \cdot p_1 + 
(1 + r^2 - 4 x_1 + 2 x_3 (1-r^2)) \epsilon \cdot p_3 ] \; , \nonumber \\
I_q^{E} &=& \int_0^1 du \int^{1-u}_0 dv (-)\, \frac{M_B^2}{\overline{M}_E^2} 
\left ( u + v (1-x_1))(1-u-v (1-x_1))(1-u-v+(1-v)(-2 x_1+ x_3) \right . 
\nonumber  \\
& & + \left .  r^2 (1-u-v-x_3 (1-v)) \right )
[ (1-r^2) \epsilon \cdot p_1 - \epsilon \cdot p_3] \; , \\ 
H_q^{E} &=& \int^1_0 du \int^{1-u}_0 dv \, \frac{m_q^2}{\overline{M}_E^2}
[ (1-r^2)(u + v (1-x_1))\epsilon 
\cdot p_1 \nonumber \\
& & + (1-2 u -2 v - 2 x_1 (1-v) + x_3 (1-r^2))\epsilon \cdot p_3] \; , \\
G_q^{E} &=& \int^1_0 du \int^{1-u}_0 dv \, \ln \frac{\overline{M}_E^2}{\mu^2}
[ (1-r^2)(1-2u-2v(1-x_1))\epsilon \cdot p_1 \nonumber \\
& & - (1-u-v-2 x_1(1-v) + x_3(1-v) +
r^2(1-u-v-x_3 (1-v)) )\epsilon \cdot p_3 ]\; ,  
\Eeqa
and 
\Beqa
C^F &=& (-) \frac{1}{6} [ (1-r^2)(1-2 x_1) \epsilon \cdot p_1 + 
2 (1 + r^2 - 4 x_1 + 2 x_3 (1-r^2)) \epsilon \cdot p_3 ] \; , \nonumber \\
I_q^{F} &=& \int^1_0 du \int^{1-u}_0 dv \, \frac{M_B^2}{\overline{M}_F^2}
\nonumber \\
& & \times \left \{ (1-r^2) (-u x_1 + v (1-x_1)) [
( v (1-x_1) + x_1 (1-u)) ((1-v)(1-x_1) + u x_1)\epsilon \cdot p_1
\right . \nonumber \\
& & + \left . ( (v (1-v) + x_1 (1-u-3 v + 2 v (u+v)))(1-r^2) -
2 ((1-u)(1-u-2 v) + v^2)x_1 x_3 \right . \nonumber \\ 
& & + \left .(2 (u-v) + (u+v)^2 -
( (1-u)(1-u-2 v) + v^2) r^2) x_3^2 )\epsilon \cdot p_3 ] \right \} \; , \\
H_q^{F} &=& \int^1_0 du \int^{1-u}_0 dv \, \frac{m_q^2}{\overline{M}_F^2}
[ (1-r^2) x_1 \epsilon \cdot p_1 - ( (v(1-x_3) + x_3 (1-u))(1-r^2) + r^2 )
\epsilon \cdot p_3 ] \; , \\
G_q^{F} &=& \int^1_0 du \int^{1-u}_0 dv \, \ln \frac{\overline{M}_F^2}{\mu^2}
\left \{ (1-r^2)(v -x_1(u+v))\epsilon \cdot p_1 \right .  \nonumber \\
& & + \left . [ 2 (x_1-x_3)(1-u-v) - r^2 (1-2 v-2 x_3(1-u-v)) ] 
\epsilon \cdot p_3 \right \}\; .  
\Eeqa

The functions $\overline{M}_E^2$ and $\overline{M}_F^2$ depend on the quark 
mass in the loop $m_q^2$ and are given by 
\Beqa
\overline{M}_E^2 &=& m_q^2 - M_B^2 [ u(1-u)r^2 + v (1-v) (1-x_1)(
-(x_1-x_3) + r^2(1-x_3)) \nonumber \\
& & - u v ( -(x_1-x_3)+r^2(2-x_1-x_3) ) ]
\Eeqa
and
\Beqa
\overline{M}_F^2 &=& m_q^2 - M_B^2 \left [ u(1-u)x_1(x_1-x_3+r^2 x_3) +
v(1-v) (1-x_1) (-(x_1-x_3)+r^2(1-x_3)) \right .\nonumber \\
&-& \left . uv ((x_1-x_3)(1-2 x_1) + 
r^2 (x_1+x_3-2 x_1 x_3)) \right ] \; . 
\Eeqa

\narrowtext

\end{document}